\documentclass[english]{emulateapj}

\usepackage[T1]{fontenc}
\usepackage[latin1]{inputenc}
\usepackage{amssymb}
\usepackage{float}
\usepackage{graphicx}
\usepackage{latexsym}
\usepackage{ae,aecompl}
\usepackage{apjfonts}

\begin{document}

\title{Correspondence principle between spherical and Euclidean wavelets}

\author{Y. Wiaux\altaffilmark{1}}
\altaffiltext{1}{Email: yves.wiaux@epfl.ch}
\affil{Signal Processing Institute, Ecole Polytechnique Fédérale de Lausanne (EPFL), CH-1015 Lausanne, Switzerland}

\author{L. Jacques}
\affil{Institut de Physique Théorique, Université catholique de Louvain (UCL), B-1348 Louvain-la-Neuve, Belgium}

\and

\author{P. Vandergheynst}
\affil{Signal Processing Institute, Ecole Polytechnique Fédérale de Lausanne (EPFL), CH-1015 Lausanne, Switzerland}

\begin{abstract}
Wavelets on the sphere are reintroduced and further developed independently
of the original group theoretic formalism, in an equivalent, but more
straightforward approach. These developments are motivated by the
interest of the scale-space analysis of the cosmic microwave background
(CMB) anisotropies on the sky. A new, self-consistent, and practical
approach to the wavelet filtering on the sphere is developed. It is
also established that the inverse stereographic projection of a wavelet
on the plane (\emph{i.e.} Euclidean wavelet) leads to a wavelet on
the sphere (\emph{i.e.} spherical wavelet). This new correspondence
principle simplifies the construction of wavelets on the sphere and
allows to transfer onto the sphere properties of wavelets on the plane.
In that regard, we define and develop the notions of directionality
and steerability of filters on the sphere. In the context of the CMB
analysis, these notions are important tools for the identification
of local directional features in the wavelet coefficients of the signal,
and for their interpretation as possible signatures of non-gaussianity,
statistical anisotropy, or foreground emission. But the generic results
exposed may find numerous applications beyond cosmology and astrophysics.
\end{abstract}


\section{Introduction}

The last decade has recognized the cosmic microwave background (CMB)
as a unique laboratory for achieving precision cosmology. The cosmological
parameters defining the structure, the energy content, and the evolution
of the universe are now determined with an impressive precision \cite{CMBpage,CMBspergel,CMBbouchet}.
However, the theoretical hypotheses on which the corresponding concordance
cosmological model relies still must be fully investigated. They notably
extend from the cosmological principle of homogeneity and isotropy,
to the models of inflation for the physics of the early universe \cite{ATwandelt,ATbartolo},
or the theory of gravitation itself, namely general relativity \cite{ATboucher1,ATboucher2}.

In the context of the concordance cosmological model, the cosmic background
radiation is understood as a unique realization of a gaussian and
stationary random signal on the sphere, arising from quantum energy
density perturbations developed in a primordial inflationary era of
the universe. In this respect, the analysis of the CMB anisotropies
on the sky is essentially confined to the study of its temperature
(and polarization) angular power spectrum.

But questioning the basic hypotheses of inflation, or of the cosmological
principle notably amounts to raise the questions of the gaussianity
and statistical isotropy (stationarity) of the CMB. New methods of
analysis of the CMB must therefore be considered. Notably, in addition to
the information on scales given through a pure
spherical harmonics decomposition, the scale-space
analysis is essential to allow the localization of features on the
sky. Such local features might for example
be associated with non-gaussianity or non-stationarity of the statistical
distribution from which the CMB arises, or with foreground emission
such as point sources. First analyses on the one-year data of the
ongoing Wilkinson Microwave Anisotropy Probe (WMAP) satellite mission
suggest a departure from both statistical isotropy and gaussianity
of the signal. The various methods applied extend from the analysis
of phase correlations \cite{ATcoles}, N-point correlation functions
\cite{SAeriksen1,SAeriksen2}, bipolar power spectra \cite{SAhajian1,SAhajian2,SAhajian3},
or local power spectra \cite{SAhansen1,SAhansen2,SAhansen3}, to multipole
vectors \cite{SAcopi,SAkatz,SAlachiezerey}. In addition, the efficiency
of the wavelet signal processing for detecting non-gaussianities in
the CMB signal was also recently established
\cite{WflatNGhobson,WflatNGbarreiro,SPmartinez,WflatNGstarck}.
While no evidence for non-gaussianity was found through wavelet analyses
in the former COsmic Background Explorer (COBE) satellite mission
data \cite{WNGbarreiro,WNGcayon}, various spherical wavelet analyses
of the one-year WMAP data also suggest a departure from non-gaussianity
or statistical isotropy \cite{WNGvielva,WNGmukherjee,WNGmcewen,WNGcruz}.
Other works also explicitly established the efficiency of the scale-space
wavelet processing for point sources detection in the foreground,
or for technical purposes such as denoising and deconvolution
\cite{WflatDsanz,WPStenorio,WflatPSvielva,WPSvielva,WflatDmaisinger}.
But a huge amount of work is still needed in this context, notably
for the identification of not only the position, but also the precise
direction, and possibly the morphology of the observed features, and
for their interpretation (see \cite{WNGmcewen} for a first approach
to the directional wavelet analysis).

In this context, a new approach to the formalism of spherical wavelets
is developed here. In § \ref{sec:Wavelets-plane}, we briefly review
the formalism for the construction of Euclidean wavelets. In § \ref{sec:Wavelets-sphere},
we reintroduce spherical wavelets independently of the original group
theoretic formalism, in a new, equivalent, and self-consistent approach.
We adopt a practical philosophy, considering wavelets as localized
filters which enable scale-space analysis, and offer an explicit reconstruction
formula for the signal considered, from its wavelet coefficients.
We also prove that the fundamental operation of dilation on the sphere
is uniquely determined from the requirement of basic natural properties.
This establishes the uniqueness of the wavelet formalism considered.
Related technical proofs are postponed to appendix \ref{sec:Wavelets-sphere-app}.
In § \ref{sec:Correspondence-principle}, we prove that the inverse
stereographic projection of a wavelet on the plane gives a wavelet
on the sphere. It is also established that the stereographic projection
is the unique projection through which this correspondence principle
holds. This new principle simplifies the construction of wavelets
on the sphere and allows to transfer wavelet properties from the plane
onto the sphere. Notice that it has been misleadingly suggested in
the original approach \cite{WSantoine3}, while not established. The
related technical proofs are detailed in appendix \ref{sec:Correspondence-principle-app}.
In § \ref{sec:Directionality-and-steerability}, the concepts of filter directionality
and steerability on the sphere are defined from the corresponding notions on the
plane. In the context of the wavelet formalism, the correspondence principle
enables to transfer these properties from wavelets on the plane to wavelets on the sphere.
The property of filter steerability allows the computation
of the rotation of a filter on itself in any direction from a simple
finite linear combination of basis filters. We also study the angular band
limitation of steerable filters and explicitly treat the examples of
steerable wavelets on the sphere defined as inverse stereographic projection 
of the derivatives of radial functions on the plane.
In § \ref{sec:CMB-local-directional}, we discuss the interest
of directional and steerable filters for the identification and interpretation
of local directional features in the context of the CMB analysis.
A numerical example illustrates our discussion. In § \ref{sec:Conclusion},
we briefly conclude.

\section{Wavelets on the plane}

\label{sec:Wavelets-plane}In this section we briefly sketch the well-known
formalism of wavelets on the plane.

On the plane as well as on the line, the notion of wavelet transform
of a signal is a powerful method of signal decomposition \cite{Wtorresani,Wmallat,Wantoine}.
We consider here a practical approach for the definition of wavelets
on the plane, which will easily be translated on the sphere. A {}``mother
wavelet'' $\psi(\vec{x})$ is first defined as a localized function
on the plane, on which affine transformations may be applied: translations,
rotations, and dilations. Second, the wavelet transform of a signal
on the plane is defined as the correlation of the signal with the
dilated and rotated versions of the mother wavelet, leading to wavelet
coefficients. This explicitly defines the scale-space nature of the
decomposition. In this context, an admissibility condition is finally
imposed on the mother wavelet by explicitly requiring the exact reconstruction
formula of the signal from its wavelet coefficients. Notice for completeness
that, in terms of the original group theoretic approach, the wavelet
decomposition is defined by the construction of the coherent states
of the group of affine transformations (translations, rotations and
dilations) on the plane. In this context, a wavelet must satisfy an
admissibility condition which ensures the square-integrability of
the unitary and irreducible representations of that group on the Hilbert
space of square-integrable functions in which the signals are defined.
This square-integrability implies that the family of wavelets obtained
by affine transformations from a mother wavelet constitutes an over-complete
frame in the considered Hilbert space, and that an exact reconstruction
formula of a signal in terms of its wavelet coefficients may be obtained.
Our more practical considerations lead identically to the same wavelet
formalism.

First, the affine transformations are defined as follows. Let us consider
the Hilbert space of square-integrable functions on the plane: $g(\vec{x})$
in $L^{2}(\mathbb{R}^{2},d^{2}\vec{x})$. In the coordinate system 
$(o,o\hat{x},o\hat{y})$, the point $\vec{x}$ on
the plane is given in Cartesian coordinates as $\vec{x}=(x,y)$, and
in polar coordinates as $\vec{x}=(r,\varphi)$. The invariant measure
on the plane, relative to the canonical
Euclidean metric in $\mathbb{R}^{2}$ is simply $d^{2}\vec{x}=dxdy$.
Generically, the action of an operator on a function in $L^{2}(\mathbb{R}^{2},d^{2}\vec{x})$
is defined by the action of the inverse of the corresponding operator
on $\mathbb{R}^{2}$, applied to the function's argument. The operator
$t(\vec{x}_{0})$ in $L^{2}(\mathbb{R}^{2},d^{2}\vec{x})$ for a translation
by an amplitude $\vec{x}_{0}=(x_{0},y_{0})$ is defined in terms of
the inverse of the translation $t_{\vec{x}_{0}}$ on points in $\mathbb{R}^{2}$.
Its action reads:
\begin{equation}
\left[t\left(\vec{x}_{0}\right)g\right]\left(\vec{x}\right)=g\left(t_{\vec{x}_{0}}^{-1}\vec{x}\right),\label{eq:wp-1}
\end{equation}
with $t_{\vec{x}_{0}}(x,y)=(x+x_{0},y+y_{0})$. The rotation operator
around the origin of coordinates $r(\chi)$ in $L^{2}(\mathbb{R}^{2},d^{2}\vec{x})$,
rotation of the wavelet around itself by an angle $\chi\in[0,2\pi[$,
is also given in terms of the inverse of the rotation $r_{\chi}$
on points in $\mathbb{R}^{2}$. It reads
\begin{equation}
\left[r\left(\chi\right)g\right]\left(\vec{x}\right)=g\left(r_{\chi}^{-1}\vec{x}\right),\label{eq:wp-2}
\end{equation}
where $r_{\chi}(x,y)$ follows from the action of the two-dimensional
rotation matrix $r_{\chi}$ on the Cartesian coordinates $(x,y)$,
or equivalently in polar coordinates $r_{\chi}(r,\varphi)=(r,\varphi+\chi)$.
The dilation operator $d(a)$ on functions in $L^{2}(\mathbb{R}^{2},d^{2}\vec{x})$
with a dilation factor $a\in\mathbb{R}_{+}^{*}$ is again defined,
in terms of the inverse of the corresponding dilation $d_{a}$ on
points in $\mathbb{R}^{2}$. It reads
\begin{equation}
\left[d\left(a\right)g\right]\left(\vec{x}\right)=a^{-1}g\left(d_{a}^{-1}\vec{x}\right),\label{eq:wp-3}
\end{equation}
with $d_{a}(x,y)=(ax,ay)$ in Cartesian coordinates, or equivalently
$d_{a}(r,\varphi)=(r_{a}(r),\varphi)$ for $r_{a}(r)=ar$ in polar
coordinates. The dilation operator on the plane is uniquely defined
(see appendix \ref{sec:Wavelets-sphere-app}) by the requirement of
the following properties. The operator $d_{a}$ on the points on $\mathbb{R}^{2}$
must be a radial (\emph{i.e.} only affecting the radial variable $r$
independently of $\varphi$, and leaving $\varphi$ invariant) and
conformal (\emph{i.e.} preserving the measure of angles in the tangent
plane at each point of $\mathbb{R}^{2}$) diffeomorphism (\emph{i.e.}
a continuously differentiable bijection on $\mathbb{R}^{2}$).
The normalization factor $a^{-1}$ in (\ref{eq:wp-3}) is also uniquely
determined by the requirement that the dilation $d(a)$ of functions
in $L^{2}(\mathbb{R}^{2},d^{2}\vec{x})$ be a unitary operator (\emph{i.e.}
preserving the scalar product in $L^{2}(\mathbb{R}^{2},d^{2}\vec{x})$,
and specifically the norm of functions).

Second, the analysis of signals goes as follows. The wavelet transform
of a signal $f(\vec{x})$ with the wavelet $\psi(\vec{x})$, localized
analysis function in $L^{2}(\mathbb{R}^{2},d^{2}\vec{x})$, is defined
as the correlation between the signal $f(\vec{x})$ and the dilated
and rotated wavelet $\psi_{\chi,a}=r(\chi)d(a)\psi$, that is as the
following scalar product:
\begin{eqnarray}
W_{\psi}^{f}\left(\vec{x}_{0},\chi,a\right) & = & \int_{\mathbb{R}^{2}}d^{2}\vec{x}\,\psi_{\vec{x}_{0},\chi,a}^{*}\left(\vec{x}\right)f\left(\vec{x}\right)\nonumber \\
& = & \langle\psi_{\vec{x}_{0},\chi,a}|f\rangle,\label{eq:wp-4}
\end{eqnarray}
 for $\psi_{\vec{x}_{0},\chi,a}=t(\vec{x}_{0})\psi_{\chi,a}$. The
wavelet coefficients $W_{\psi}^{f}(\vec{x}_{0},\chi,a)$ represent
the characteristics of the signal for each analysis scale $a$, direction
$\chi$, and position $\vec{x}_{0}$.

Finally, the synthesis of a signal $f(\vec{x})$ from its wavelet
coefficients reads:

\begin{eqnarray}
f\left(\vec{x}\right) & = & \int_{0}^{2\pi}d\chi\int_{0}^{+\infty}\frac{da}{a^{3}}\int_{\mathbb{R}^{2}}d^{2}\vec{x}_{0}\nonumber \\
&  & W_{\psi}^{f}(\vec{x}_{0},\chi,a)\left[t\left(\vec{x}_{0}\right)r\left(\chi\right)L_{\psi}\psi_{a}\right]\left(\vec{x}\right).\label{eq:wp-5}
\end{eqnarray}
 In this relation, the operator $L_{\psi}$ in $L^{2}(\mathbb{R}^{2},d^{2}\vec{x})$
is defined by the simple division by a constant: $[L_{\psi}g](\vec{x})=g(\vec{x})/C_{\psi}$.
This exact reconstruction formula holds if and only if the wavelet
$\psi(\vec{x})$ satisfies the following admissibility condition:
\begin{equation}
0<C_{\psi}=\int_{\mathbb{R}^{2}}d^{2}\vec{k}\,\frac{|\widehat{\psi}\big(\vec{k}\big)|^{2}}{|\vec{k}|^{2}}<\infty,\label{eq:wp-6}
\end{equation}
with the normalization convention $\hat{\psi}(\vec{k})=\int d\vec{x}\, e^{-i\vec{k}\cdot\vec{x}}\psi(\vec{x})$
for the Fourier transform of functions in the plane. The zero-mean
is therefore a necessary condition for the wavelet admissibility in
$L^{2}(\mathbb{R}^{2},d^{2}\vec{x})$:
\begin{equation}
\int_{\mathbb{R}^{2}}d^{2}\vec{x}\,\psi\left(\vec{x}\right)=0.\label{eq:wp-7}
\end{equation}
It is also well-known that under the additional requirement that
$\psi(\vec{x})$ be in $L^{1}\cap L^{2}(\mathbb{R}^{2},d^{2}\vec{x})$,
the zero-mean condition (\ref{eq:wp-7}) implies the exact admissibility
condition (\ref{eq:wp-6}). Wavelets on the plane may therefore be
easily built.

\section{Wavelets on the sphere}

\label{sec:Wavelets-sphere}The formalism of wavelets on the sphere
was originally established in a group theoretic framework \cite{WSantoine2,WSantoine1,WSantoine3,WSdemanet,WSbogdanova}.
In this section we reintroduce the notion of wavelets on the sphere
through a new and completely self-consistent approach. The resulting
formalism is equivalent to the original one, but more practical and
straightforward. The structure of our approach follows, in perfect
analogy with the formalism of wavelets on the plane introduced in
the former section.

For the clarity of the expressions, we denote functions and operators
on the sphere in uppercase letters, by opposition to the lowercase
letters denoting functions and operators on the plane. Identically
to our approach in the plane, a {}``mother wavelet'' $\Psi(\omega)$
is first defined as a localized function on the unit sphere, on which
affine transformations may be applied: translations, rotations, and
dilations. Second, the wavelet transform of a signal on the sphere
is defined as the correlation of the signal with the dilated and rotated
versions of the mother wavelet, leading to wavelet coefficients, defining
the scale-space nature of the decomposition on the sphere. Third,
an admissibility condition is imposed on the mother wavelet by explicitly
requiring the exact reconstruction formula of the signal from its
wavelet coefficients. We again notice that, similarly to the formalism
in the plane, in a group theoretic approach, the wavelet decomposition
is defined by the construction of generalized coherent states for
the affine transformations on the sphere (translations, rotations
and dilations) contained in the conformal group of the sphere, $SO(1,3)$.
As already stated, our approach leads to the same final wavelet formalism.

First, the affine transformations are defined as follows on square-integrable
functions $G(\omega)$ in $L^{2}(S^{2},d\Omega)$ on the unit sphere.
The point $\omega$ on the sphere is given in spherical coordinates
as $\omega=(\theta,\varphi)$. In an orthonormal Cartesian coordinate
system $(o,o\hat{x},o\hat{y},o\hat{z})$ centered on the unit sphere,
the polar angle, or co-latitude, $\theta\in[0,\pi]$ represents the
angle between the vector identifying $\omega$ and the axis $o\hat{z}$.
The azimuthal, or longitudinal, angle $\varphi\in[0,2\pi[$, not defined
though for $\theta\in\{0,\pi\}$, represents the angle between the
projection of this vector in the plane $(o,o\hat{x},o\hat{y})$ and
the axis $o\hat{x}$. The invariant measure on the sphere,
relative to the canonical metric on $S^{2}$ induced from the Euclidean
metric in three dimensions is $d\Omega=d\cos\theta d\varphi$. Once
again, generically, the action of an operator on a function in $L^{2}(S^{2},d\Omega)$
is defined by the action of the inverse of the corresponding operator
on $S^{2}$, applied to the function's argument.

The action of a rotation $\rho\in SO(3)$ in three dimensions on a
function $G$ on the sphere, is defined by the operator $R(\rho):G(\omega)\rightarrow[R(\rho)G](\omega)=G(R_{\rho}^{-1}\omega)$
in $L^{2}(S^{2},d\Omega)$. The operator $R_{\rho}$ may be decomposed
in three consecutive rotations, respectively around the axes of coordinates
$o\hat{z}$, $o\hat{y}$, and $o\hat{z}$, and defined by the Euler
angles $(\varphi_{0},\theta_{0},\chi)$, with $\theta_{0}\in[0,\pi]$
and $\varphi_{0},\chi\in[0,2\pi[$: $R_{\rho}=R_{\varphi_{0},\theta_{0},\chi}=R_{\varphi_{0}}^{\hat{z}}R_{\theta_{0}}^{\hat{y}}R_{\chi}^{\hat{z}}$.
The inverse rotation $R_{\rho}^{-1}$ is characterized by opposite
Euler angles in the reverse order: $R_{\varphi_{0},\theta_{0},\chi}^{-1}=R_{-\chi,-\theta_{0},-\varphi_{0}}=R_{\pi-\chi,\theta_{0},\pi-\varphi_{0}}$.
In the particular context of the analysis of functions on the sphere,
the variable $(\varphi_{0},\theta_{0},\chi)$ in the parameter space
of the group $SO(3)$ may decomposed as $(\omega_{0},\chi)$ in $S^{2}\otimes[0,2\pi[$,
where $\omega_{0}$ defines a position on the sphere $S^{2}$, and
$\chi$ a direction in $[0,2\pi[$ at each point. In that context,
the rotation operator on a function $G$ on the sphere is decomposed
as $R(\rho)=R(\omega_{0})R^{\hat{z}}(\chi)$, where $R(\omega_{0})=R^{\hat{z}}(\varphi_{0})R^{\hat{y}}(\theta_{0})$
defines the motion or translation of the function by $\omega_{0}$,
and $R^{\hat{z}}(\chi)$ defines its rotation by $\chi$ on itself.
The operator $R(\omega_{0})$ on the sphere for a motion of amplitude
$\omega_{0}$ is defined in terms of the inverse of the motion
operator $R_{\omega_{0}}$ on points in $S^{2}$. It reads
\begin{equation}
\left[R\left(\omega_{0}\right)G\right]\left(\omega\right)=G\left(R_{\omega_{0}}^{-1}\omega\right),\label{eq:ws-1}
\end{equation}
where $R_{\omega_{0}}(\theta,\varphi)$ readily follows from the
action of the three-dimensional rotation matrices $R_{\theta_{0}}^{\hat{y}}$
and $R_{\varphi_{0}}^{\hat{z}}$ on the Cartesian coordinates in three
dimensions associated with the point $\omega$ on the sphere. The
rotation operator $R^{\hat{z}}(\chi)$ of a function around itself,
by an angle $\chi\in[0,2\pi[$, follows again from the inverse of
the rotation operator $R_{\chi}^{\hat{z}}$ on points in $S^{2}$.
It is given as
\begin{equation}
\left[R^{\hat{z}}\left(\chi\right)G\right]\left(\omega\right)=G\left({R_{\chi}^{\hat{z}}}^{-1}\omega\right),\label{eq:ws-2}
\end{equation}
where $R_{\chi}^{\hat{z}}(\theta,\varphi)=(\theta,\varphi+\chi)$
also follows from the action of the three-dimensional rotation matrix
$R_{\chi}^{\hat{z}}$ on the Cartesian coordinates in three dimensions
associated with the point $\omega$.

The dilation operator $D(a)$ on functions in $L^{2}(S^{2},d\Omega)$,
for a dilation factor $a\in\mathbb{R}_{+}^{*}$, is defined in terms
of the inverse of the corresponding dilation $D_{a}$ on points in
$S^{2}$ as
\begin{equation}
\left[D\left(a\right)G\right]\left(\omega\right)=\lambda^{1/2}\left(a,\theta\right)G\left(D_{a}^{-1}\omega\right).\label{eq:ws-3}
\end{equation}
In spherical coordinates, the dilated point is given by $D_{a}(\theta,\varphi)=(\theta_{a}(\theta),\varphi)$
with $\theta_{a}(\theta)=2\arctan(a\tan(\theta/2))$. This corresponds
to a linear dilation of $\tan(\theta/2)$: $\tan(\theta_{a}(\theta)/2)=a\tan(\theta/2)$.
The dilation operator therefore maps the sphere without its South pole
on itself: $\theta_a(\theta):\theta\in [0,\pi[\rightarrow\theta_a\in [0,\pi[$.
It is established in appendix \ref{sec:Wavelets-sphere-app} that
this dilation operator on the sphere is uniquely defined by the requirement
of the same basic and natural properties as on the plane. The dilation
$D_{a}$ of points on $S^{2}$ must be a radial (\emph{i.e.} only
affecting the radial variable $\theta$ independently of $\varphi$,
and leaving $\varphi$ invariant) and conformal (\emph{i.e.} preserving
the measure of angles in the tangent plane at each point of $S^{2}$)
diffeomorphism (\emph{i.e.} a continuously differentiable bijection
on $S^{2}$). The conformal factor $\lambda(a,\theta)$
reads $\lambda^{1/2}(a,\theta)=a^{-1}[1+\tan^{2}(\theta/2)]/[1+a^{-2}\tan^{2}(\theta/2)]$,
also equal to $\lambda^{1/2}(a,\theta)=2a/[(a^{2}-1)\cos\theta+(a^{2}+1)]$
as defined in the original formalism \cite{WSantoine2}. The normalization
by $\lambda^{1/2}(a,\theta)$ in (\ref{eq:ws-3}) is uniquely determined
by the requirement that the dilation $D(a)$ of functions in $L^{2}(S^{2},d\Omega)$
be a unitary operator (\emph{i.e.} preserving the scalar product in
$L^{2}(S^{2},d\Omega)$, and specifically the norm of functions).
Notice that in the limit $\theta\rightarrow0$, this dilation factor
on the sphere naturally reduces to first order in $\theta$ to the
normalization constant on the plane: $\lambda^{1/2}(a,\theta)\rightarrow a^{-1}$.

Second, the analysis of signals goes as follows. The wavelet transform
$W_{\Psi}^{F}(\omega_{0},\chi,a)$ of a signal $F(\omega)$ with the
wavelet $\Psi(\omega)$, localized analysis function in $L^{2}(S^{2},d\Omega)$
on the sphere, is defined as the correlation between $F(\omega)$
and the dilated and rotated wavelet $\Psi_{\chi,a}=R^{\hat{z}}(\chi)D(a)\Psi$,
that is again as the scalar product:
\begin{eqnarray}
W_{\Psi}^{F}\left(\omega_{0},\chi,a\right) & = & \int_{S^{2}}d\Omega\,\Psi_{\omega_{0},\chi,a}^{*}\left(\omega\right)F\left(\omega\right)\nonumber \\
& = & \langle\Psi_{\omega_{0},\chi,a}|F\rangle,\label{eq:ws-4}
\end{eqnarray}
with $\Psi_{\omega_{0},\chi,a}=R(\omega_{0})\Psi_{\chi,a}$. The
wavelet coefficients $W_{\Psi}^{F}(\omega_{0},\chi,a)$ represent
the characteristics of the signal for each analysis scale $a$, direction
$\chi$, and position $\omega_{0}$.

Third, the synthesis of a signal $F(\omega)$ from its wavelet coefficients
reads as:
\begin{eqnarray}
F\left(\omega\right) & = & \int_{0}^{2\pi}d\chi\int_{0}^{+\infty}\frac{da}{a^{3}}\int_{S^{2}}d\Omega_{0}\nonumber \\
&  & W_{\Psi}^{F}\left(\omega_{0},\chi,a\right)\left[R\left(\omega_{0},\chi\right)L_{\Psi}\Psi_{a}\right]\left(\omega\right).\label{eq:ws-5}
\end{eqnarray}
In this relation, the operator $L_{\Psi}$ in $L^{2}(S^{2},d\Omega)$
is defined by the following action on the spherical harmonics coefficients
of functions: $\widehat{L_{\Psi}G}_{lm}=\widehat{G}_{lm}/C_{\Psi}^{l}$,
with $l\in\mathbb{N}$, $m\in\mathbb{Z}$, and $|m|\leq l$. On the
contrary to $L_{\psi}$ on the plane, the operator $L_{\Psi}$ does
not summarize to the division by a constant. Notice that the \emph{a
priori} arbitrary choice of scale integration measure $da/a^{3}$
is uniquely fixed by requiring the correspondence principle discussed
in the next section. In appendix \ref{sec:Wavelets-sphere-app}, a
detailed proof shows that this exact reconstruction formula holds
if and only if the spherical harmonics transform $\widehat{\Psi}_{lm}$
of the wavelet $\Psi(\omega)$ satisfies the following admissibility
condition:
\begin{equation}
0<C_{\Psi}^{l}=\frac{8\pi^2}{2l+1}\sum_{|m|\leq l}\int_{0}^{+\infty}\frac{da}{a^{3}}\,|\widehat{\left(\Psi_{a}\right)}_{lm}|^{2}<\infty,\label{eq:ws-6}
\end{equation}
for all $l\in\mathbb{N}$. This condition is slightly less restrictive than the admissibility condition required in the original group theoretic formalism for the square-integrability of the group representation considered, and  given in (\ref{eq:a-cp-4}). It may also be shown that the following relation defines a necessary condition for the wavelet admissibility
in $L^{2}(S^{2},d\Omega)$ \cite{WSantoine2}:
\begin{equation}
\int_{S^{2}}d\Omega\,\frac{\Psi\left(\theta,\varphi\right)}{1+\cos\theta}=0.\label{eq:ws-7}
\end{equation}

Finally, in the limit $\theta\rightarrow0$, the formalism of spherical
wavelets simply identifies with the formalism of Euclidean wavelets.
This Euclidean limit \cite{WSantoine2} is naturally established by
direct identification of the relations (\ref{eq:wp-1}) to (\ref{eq:wp-7})
on the plane, one by one with the relations (\ref{eq:ws-1}) to (\ref{eq:ws-7})
on the sphere, to first order in $\theta$. In particular, the identification
of the admissibility conditions (\ref{eq:wp-6}) and (\ref{eq:ws-6})
is made obvious in terms of their reformulation, as expressed in appendix
\ref{sec:Correspondence-principle-app}.

Also notice that the wavelet formalism may be similarly defined in
the space of integrable functions, on the plane, $L^{1}(\mathbb{R}^{2},d^{2}\vec{x})$,
as on the sphere, $L^{1}(S^{2},d\Omega)$. The dilation operator on
functions must be redefined consistently in such a way that it still
preserves the norm of the functions on which it is applied. The measure
of integration on scales changes and the admissibility condition on
the sphere is modified consistently. In that formalism, the reconstruction
formula contains an additional low frequency term, and the corresponding
necessary admissibility condition on the sphere reduces to an exact
zero-mean condition:$\int_{S^{2}}d\Omega\,\Psi(\theta,\varphi)=0$
\cite{WSantoine3}. But we do not develop this formalism here.

\section{Correspondence principle}

\label{sec:Correspondence-principle}In the previous sections, we
established the wavelet formalism, independently on the plane and
on the sphere. Wavelets on the plane are well-known, and may be easily
defined in terms of the zero-mean condition (\ref{eq:wp-7}) for a
function both integrable and square-integrable, implying the admissibility
condition (\ref{eq:wp-6}). On the contrary, the admissibility condition
(\ref{eq:ws-6}) for the definition of wavelets on the sphere is difficult
to check in practice. In this section, we prove that the inverse stereographic
projection of a wavelet on the plane leads to a wavelet on the sphere.
For clarity, the related technical proofs are postponed to appendix
\ref{sec:Correspondence-principle-app}. Beyond its pure theoretical
interest, this new correspondence principle between the wavelet formalisms
on the plane and on the sphere is of great practical use. Indeed,
it enables the construction of wavelets on the sphere by simple projection
of wavelets on the plane. In that respect, it also naturally allows
to transfer wavelet properties from the plane onto the sphere, as illustrated
in the next section.

First, appendix \ref{sec:Correspondence-principle-app} establishes
that the projection of a wavelet on the plane leads to a wavelet on
the sphere under specific requirements on the corresponding projection
operator. That is to say, if the function $\psi(r,\varphi)$ in $L^{2}(\mathbb{R}^{2},d^{2}\vec{x})$
satisfies the wavelet admissibility condition (\ref{eq:wp-6}) on
the plane, then the function
\begin{equation}
\Psi\left(\theta,\varphi\right)=\left[\Pi^{-1}\psi\right]\left(\theta,\varphi\right),\label{eq:cp-1}
\end{equation}
in $L^{2}(S^{2},d\Omega)$ satisfies the wavelet admissibility condition
(\ref{eq:ws-6}) on the sphere. The projection operator $\Pi$ between
functions in $L^{2}(S^{2},d\Omega)$ and in $L^{2}(\mathbb{R}^{2},d^{2}\vec{x})$
on the two manifolds is defined in terms of the inverse of the corresponding
projection operator $\pi$ between points on the sphere $S^{2}$ and
on the plane $\mathbb{R}^{2}$, applied to the argument of the function
considered. The proof of this correspondence principle relies on the
requirement that the projection operator be a unitary, radial, and
conformal diffeomorphism. Let us recall that the unit sphere on which
spherical wavelets are defined is centered at the origin of the orthonormal
Cartesian coordinate system $(o,o\hat{x},o\hat{y},o\hat{z})$ in three
dimensions, with spherical coordinates $(\theta,\varphi)$. For simplicity,
and without loss of generality, we consider a geometrical setting
where the plane on which the Euclidean wavelets are defined is parallel
to the plane $(o\hat{x},o\hat{y})$, at an arbitrary height $z_{0}$.
The polar coordinates $(r,\varphi)$ on the plane are defined relatively
to the coordinate system $(o,o\hat{x},o\hat{y})$. The polar angle $\varphi$
on the plane and the azimuthal angle $\varphi$ on the sphere are
therefore identified with one another. We consider specifically, still
without loss of generality, the plane tangent to the sphere at the
North pole, $z_{0}=1$ (see Fig. \ref{cap:Stereographic-projection}
below). With this choice of coordinates, the operator $\pi$ of projection
of points from $S^{2}$ onto $\mathbb{R}^{2}$ must naturally be a
radial diffeomorphism. That is, it must be a continuously differentiable bijection
between $S^{2}$ and ${R}^{2}$,
which only relates the radial variables $r$ on the plane and $\theta$
on the sphere independently of $\varphi$, and which leaves $\varphi$ invariant.
It must also define a conformal mapping between the sphere and the
plane (\emph{i.e.} the metric induced on the plane by the transformation
$r(\theta)$ is conformally equivalent to the Euclidean metric on
the plane). This property is essential
to ensure that the measure of angles and directions (orientations) is preserved
by projection. The projection operator $\Pi$ between functions in
$L^{2}(S^{2},d\Omega)$ and in $L^{2}(\mathbb{R}^{2},d^{2}\vec{x})$
on the two manifolds must also be unitary (\emph{i.e.} preserve the
scalar product between $L^{2}(S^{2},d\Omega)$ and $L^{2}(\mathbb{R}^{2},d^{2}\vec{x})$,
and specifically the norm of functions).

Second, the correspondence principle may also be extended in the following
way. The rotation $R^{\hat{z}}(\chi)$, acting on functions on the
sphere through the azimuthal angular variable $\varphi$, is conjugate
to the rotation $r(\chi)$, acting on functions on the tangent plane
at the North pole through the polar angular variable $\varphi$, by
any radial projection:
\begin{equation}
R^{\hat{z}}\left(\chi\right)=\Pi^{-1}r\left(\chi\right)\Pi.\label{eq:cp-2}
\end{equation}
As shown in appendix \ref{sec:Correspondence-principle-app}, for
a unitary, radial and conformal projection operator $\Pi$ between
$L^{2}(S^{2},d\Omega)$ and $L^{2}(\mathbb{R}^{2},d^{2}\vec{x})$,
a conjugation relation holds between the dilation operators (\ref{eq:ws-3})
and (\ref{eq:wp-3}), through the projection $\Pi$:
\begin{equation}
D\left(a\right)=\Pi^{-1}d\left(a\right)\Pi.\label{eq:cp-3}
\end{equation}
It appears from these two conjugation relations that the operations
of dilation by $a\in\mathbb{R}_{+}^{*}$ and rotation by $\chi\in[0,2\pi[$
of a spherical wavelet defined as projection of a Euclidean wavelet
may be simply performed on the plane before inverse projection on
the sphere:
\begin{equation}
\Psi_{\chi,a}\left(\theta,\varphi\right)=\left[\Pi^{-1}\psi_{\chi,a}\right]\left(\theta,\varphi\right).\label{eq:cp-4}
\end{equation}
The dilated and rotated wavelets on the sphere may therefore be built
in an extremely straightforward way with the intuitive operations
of dilation (\ref{eq:wp-3}) and rotation (\ref{eq:wp-2}) on the
plane, forgetting the corresponding operators (\ref{eq:ws-3}) and
(\ref{eq:ws-2}) on the sphere. Also notice that, as all considered
operators are unitary, if the wavelet on the plane is normalized,
the corresponding wavelet on the sphere remains naturally normalized.
Only motions (translations) by $\omega_{0}$ have to be explicitly
performed on the sphere as the corresponding operator (\ref{eq:ws-1})
is not the conjugate of the operator (\ref{eq:wp-1}) of translation
by any $\vec{x}_{0}$ on the plane.

Third, appendix \ref{sec:Correspondence-principle-app} also establishes
that the stereographic projection is the unique radial conformal diffeomorphism
mapping the sphere $S^{2}$ onto the plane $\mathbb{R}^{2}$. The
unitary stereographic projection operator between functions $G$
in $L^{2}(S^{2},d\Omega)$ and $g$ in $L^{2}(\mathbb{R}^{2},d^{2}\vec{x})$
and its inverse respectively read
\begin{equation}
\left[\Pi G\right]\left(\vec{x}\right)=\mu^{1/2}\left(r\right)G\left(\pi^{-1}\vec{x}\right),\label{eq:cp-5}
\end{equation}
and
\begin{equation}
\left[\Pi^{-1}g\right]\left(\omega\right)=\mu^{-1/2}\left(r\left(\theta\right)\right)g\left(\pi\omega\right).\label{eq:cp-6}
\end{equation}
The radial conformal diffeomorphism between points is given as $\pi(\theta,\varphi)=(r(\theta),\varphi)$
and $\pi^{-1}(r,\varphi)=(\theta(r),\varphi)$, and for $r(\theta)=2\tan(\theta/2)$
and $\theta(r)=2\arctan(r/2)$. The diffeomorphism $r(\theta)$ and
its inverse $\theta(r)$ explicitly define the stereographic projection
of points on the sphere onto points on the plane, and its inverse.
This stereographic projection maps the sphere, without its South pole,
on the entire plane: $r(\theta):\theta\in[0,\pi[\rightarrow[0,\infty[$.
Geometrically, it projects a point $\omega=(\theta,\varphi)$ on the
unit sphere onto a point $\vec{x}=(r,\varphi)$ on the tangent plane
at the North pole, co-linear with $\omega$ and the South pole (see
Fig. \ref{cap:Stereographic-projection}). The conformal factor $\mu(r)$
is given as $\mu^{1/2}(r)=(1+(r/2)^{2})^{-1}$. The normalization
$\mu^{1/2}(r)$ is required to ensure the unitarity of the projection
operator $\Pi$ between $L^{2}(S^{2},d\Omega)$ and $L^{2}(\mathbb{R}^{2},d^{2}\vec{x})$.
Conversely, $\mu^{-1/2}(r(\theta))=1+\tan^{2}(\theta/2)$ is the conformal
factor associated with the inverse radial conformal mapping $\pi^{-1}$, which ensures
the unitarity of $\Pi^{-1}$.

\begin{figure}[H]
\begin{center}\includegraphics[width=8cm]{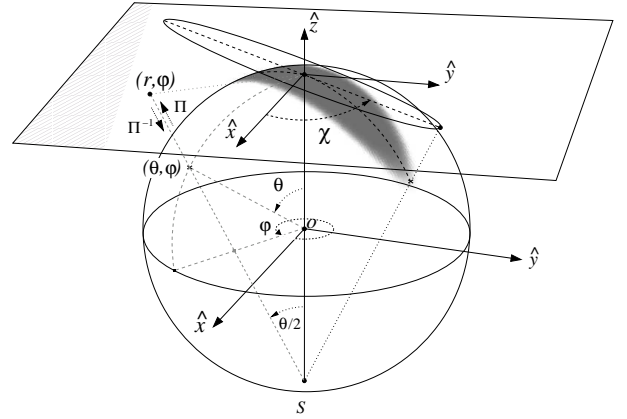}\end{center}

\caption{\label{cap:Stereographic-projection}Relation established between
points on the sphere $(\theta,\varphi)$ and on its tangent plane
at the North pole $(r,\varphi)$ through the stereographic projection
$\pi$ and its inverse $\pi^{-1}$. The same relation holds between
functions living on each of the two manifolds, as illustrated by the
shadow on the sphere and the localized region on the plane. The corresponding
operators $\Pi$ and $\Pi^{-1}$ enable the establishment of a correspondence
principle between the wavelet formalisms on the plane and on the sphere.}
\end{figure}

Let us consider one example, as illustration of the correspondence
principle through the stereographic projection. On the plane, the
derivatives of gaussians in a specific direction, say $\hat{x}$,
are well-known examples of wavelets. The normalized (negative) first
derivative of gaussian in direction $\hat{x}$ reads, after dilation
by $a$ and rotation by $\chi$: $\psi_{\chi,a}(r,\varphi)=\sqrt{2/\pi}\, re^{-r^{2}/2a^{2}}\cos(\varphi-\chi)/a^{2}$.
The corresponding wavelet on the sphere, normalized, dilated by $a$
and rotated by $\chi$, but still at the North pole, is simply obtained
by the action of the inverse stereographic projection (\ref{eq:cp-6}):
$\Psi_{\chi,a}(\theta,\varphi)=(1+\tan^{2}(\theta/2))\,\sqrt{2/\pi}\,2\tan(\theta/2)e^{-2\tan^{2}(\theta/2)/a^{2}}\cos(\varphi-\chi)/a^{2}$.
Notice that, depending on the application, the exact zero mean of
a filter is often an appreciated property, as the corresponding filtering
completely erases constant signals. In the $L^{2}$ formalism, this
property may be considered as an additional practical requirement
for the choice of the mother wavelet. As illustrated by the present
example, the stereographic projection of any odd-numbered derivative
of a radial function already satisfies this zero mean condition, in
addition to the admissibility condition (\ref{eq:ws-6}).

Finally, these developments necessitate the following remarks. Notice
on the one hand, that the conformal relation between the Euclidean
invariant measure on the plane and the measure induced from the stereographic
projection, $r(\theta)dr(\theta)d\varphi=\mu^{-1}(r(\theta))\sin\theta d\theta d\varphi$
with $r(\theta)=2\tan(\theta/2)$, readily implies the correspondence
of the necessary wavelet admissibility conditions on the plane and
on the sphere. If a function satisfies the necessary wavelet admissibility
condition (\ref{eq:wp-7}) on the plane, then its inverse stereographic
projection satisfies the necessary wavelet admissibility
condition (\ref{eq:ws-7}) on the sphere. This result gives us intuition
on the otherwise non-trivial complete demonstration of the correspondence
principle (\ref{eq:cp-1}) established in appendix \ref{sec:Correspondence-principle-app},
which links the necessary and sufficient wavelet admissibility conditions.
On the other hand, we also emphasize that, in the limit $\theta\rightarrow0$,
we get the first order relation $r(\theta)\rightarrow\theta$, which
corresponds to the identification between the considered portion of
the sphere around the North pole, with the tangent plane at that same
point. The factors of unitarity also tend to unity. The stereographic
projection summarizes to the identity operator. The wavelets on the
plane are therefore identified with their projection on the sphere,
in complete coherence with the Euclidean limit discussed in §
\ref{sec:Wavelets-sphere}.

\section{Filter directionality and steerability}

\label{sec:Directionality-and-steerability}In this section, we discuss
the notions of directionality and steerability of
filters on the plane and on the sphere. On the sphere, these developments
constitute a new advance for the scale-space analysis of signals. Their interest will
be illustrated in the next section in the context of the CMB analysis.
In a first subsection, we recall the well-known notions of directionality and steerability
on the plane. We introduce the corresponding definitions 
on the sphere. We also show that these notions are transferred from
the plane onto the sphere through the stereographic projection.
In the context of the wavelet formalism, the correspondence principle 
established in the former section therefore enables to transfer the directionality
and steerability of wavelets, from the plane onto the sphere.
In a second subsection, we characterize steerable filters in terms of their
band limitation in the Fourier index conjugate to the angular variable
$\varphi$. In the last subsection, we finally develop the examples of
steerable wavelets on the sphere defined as inverse stereographic projection of
the derivatives of radial functions on the plane.

\subsection{Definitions and correspondence}

First, we recall the notions of directionality and steerability on the plane.
Locally, at each point of coordinates $\vec{x}=(r,\varphi)$ on the plane $\mathbb{R}^{2}$, directions
are defined in the tangent plane, in terms of the rotation angle $\chi\in[0,2\pi[$.
The origin of angles ($\chi=0$) in the tangent plane is defined by the direction tangent
to the line passing through the point considered, from the origin $o$ of coordinates, and making angle
$\varphi$ with the axis $o\hat{x}$ (direction of increasing $r$) .
The directionality of a filter $g(r,\varphi)$ in $L^{2}(\mathbb{R}^{2},d^{2}\vec{x})$
on the plane may be measured through its auto-correlation function,
defined as the scalar product of the rotations of the filter in two
different directions $\chi$ and $\chi'$, and depending on the difference
$\Delta\chi=\chi-\chi'$:
\begin{equation}
C^{g}\left(\Delta\chi\right)=\langle r\left(\chi\right)g|r\left(\chi'\right)g\rangle.\label{eq:ds-1}
\end{equation}
The steerability of a directional filter $g(r,\varphi)$ is defined
in $L^{2}(\mathbb{R}^{2},d^{2}\vec{x})$ on the plane \cite{STfreeman,STsimoncelli}
by the requirement that any rotated version $r(\chi)g$ of the filter
be expressed as a linear combination of a finite number of rotations
of the filter in specific directions $\chi_{m}$. This is mathematically
defined by the relation $[r(\chi)g](r,\varphi)=\sum_{m=1}^{M}k_{m}(\chi)[r(\chi_{m})g](r,\varphi)$,
where the weights $k_{m}(\chi)$ , with $1\leq m\leq M$, and $M\in\mathbb{N}$,
are called interpolation functions. More generally, one requires that
$r(\chi)g$ be expressed in terms of a finite number of independent
basis filters $g_{m}(r,\varphi)$ in $L^{2}(\mathbb{R}^{2},d^{2}\vec{x})$,
which are not necessarily specific rotations of $g(r,\varphi)$:
\begin{equation}
\left[r\left(\chi\right)g\right]\left(r,\varphi\right)=\sum_{m=1}^{M}k_{m}\left(\chi\right)g_{m}\left(r,\varphi\right).\label{eq:ds-2}
\end{equation}

Second, we can define the notions of directionality
and steerability on the sphere in perfect analogy with their definition on the plane.
Locally, at each point of coordinates 
$\omega=(\theta,\varphi)$ on the sphere $S^{2}$, 
directions are defined in terms of the third Euler angle $\chi\in[0,2\pi[$,
which identifies the directions in the tangent plane. The origin of angles
($\chi=0$) in the tangent plane is defined by the direction tangent
to the meridian passing through the point considered (direction of increasing $\theta$).
Let us recall that the non-existence of 
differentiable vector fields on $S^{2}$ rules out the definition of directions
globally on the sphere.
The directionality of a filter $G(\theta,\varphi)$ in $L^{2}(S^{2},d\Omega)$
on the sphere is measured through its auto-correlation function,
also defined as the scalar product of the rotations of the filter in two
different directions $\chi$ and $\chi'$, and
depending on difference $\Delta\chi=\chi-\chi'$:
\begin{equation}
C^{G}\left(\Delta\chi\right)=\langle R^{\hat{z}}\left(\chi\right)G|R^{\hat{z}}\left(\chi'\right)G\rangle.\label{eq:ds-3}
\end{equation}
The steerability of a directional filter $G(\theta,\varphi)$ is defined
in $L^{2}(S^{2},d\Omega)$ on the sphere by the relation
\begin{equation}
\left[R^{\hat{z}}\left(\chi\right)G\right]\left(\theta,\varphi\right)=\sum_{m=1}^{M}k_{m}\left(\chi\right)G_{m}\left(\theta,\varphi\right).\label{eq:ds-4}
\end{equation}
Again, the weights $k_{m}(\chi)$,
with $1\leq m\leq M$, and $M\in\mathbb{N}$ are the interpolation functions.
The basis filters $G_{m}(\theta,\varphi)$ are not necessarily specific rotations of
$G(\theta,\varphi)$.

Third, we show that these properties are transferred from the plane onto the sphere through
the stereographic projection. Considering the stereographic projection $G=\Pi^{-1}g$,
the unitarity of the
operator $\Pi$ ensures that the auto-correlation function of the projected filter
is identical to the auto-correlation function of the original filter
on the plane: $C^{(\Pi^{-1}g)}(\Delta\chi)=C^{g}(\Delta\chi)$.
Moreover, if the relation of steerability (\ref{eq:ds-2}) holds for
$g(r,\varphi)$ on the plane, then the relation of steerablity
(\ref{eq:ds-4}) obviously holds
for $G=\Pi^{-1}g$ on the sphere, with the same interpolation functions
$k_{m}(\chi)$ , for $1\leq m\leq M$, and $M\in\mathbb{N}$,
and for the basis filters $G_{m}(\theta,\varphi)$ defined as $G_{m}=\Pi^{-1}g_{m}$.
This property explicitly relies on the conjugation relation (\ref{eq:cp-2})
induced by any radial projection operator. If wavelet filters are considered,
the correspondence principle established in § \ref{sec:Correspondence-principle}
therefore enables to transfer the directionality and steerability of wavelets,
from the plane onto the sphere.

We finally discuss the interest of the concepts introduced, on the plane as on the
sphere. Let us consider on the one hand the notion of filter directionality.
Any non-axisymmetric filter will be defined as directional in the
sense that its auto-correlation is not a constant function of $\Delta\chi$.
We consider as a good directional filter, a filter for which the auto-correlation,
 (\ref{eq:ds-1}) or (\ref{eq:ds-3}),
is a rapidly decreasing function of $\Delta\chi$. In this regard,
the ideal directional filter would have the expression of a $\delta$
distribution in the angle $\varphi$, in such a way that its
auto-correlation be a $\delta(\Delta\chi)$ distribution.
Notice that other definitions of directionality on the plane and on the sphere
may be found in the literature \cite{WSantoine3,WSdemanet,Wantoine}.
Our definitions have the non-negligible
advantage of corresponding to one another through the stereographic projection between
the sphere and the plane. In
these terms, the peakedness of the auto-correlation function is a
measure of the sensitivity of the filter to directions. From a practical
point of view, it also measures how the identification of directions,
in terms of the maximization of the filtering coefficients of a considered
signal, is sensitive to any kind of noise inevitably affecting the
estimation of these coefficients. The directionality is consequently
a key criterion for the choice of the filter for the identification
of local directions of a signal. Consider on the other hand the notion
of filter steerability. Through a linear filtering, if a relation of 
steerability holds for a filter, (\ref{eq:ds-2}) or (\ref{eq:ds-4}), then
the same relation holds for the filtering coefficients of the signal considered.
This is specifically true in the analysis of a signal with wavelet filters.
The steerability therefore allows the computation
of filtering coefficients of a signal in all directions at the cost
of the computation of $M$ filtering coefficients. This property may
therefore reduce the computation cost of filtering by a non-negligible
factor. Its direct interest in the perspective of the CMB analysis
is suggested in § \ref{sec:CMB-local-directional}.
In the following subsection, we discuss the band limitation in the
Fourier index conjugate to the variable $\varphi$ for directional
and steerable filters.

\subsection{Angular band limitation}

We show here that the notions of ideal directionality and steerability represent competing
concepts on the plane as on the sphere, in terms of the band
limitation of the considered filters in the Fourier index conjugate
to the angular variable $\varphi$. 
The following results are discussed on the plane,
for functions $g(r,\varphi)$ in $L^{2}(\mathbb{R}^{2},d^{2}\vec{x})$.
They may be identically read on the sphere for $G(\theta,\varphi)$
in $L^{2}(S^{2},d\Omega)$, through the substitution of $g$ by $G$.
Let us thus consider
the Fourier decomposition of the filter $g(r,\varphi)$ in the variable
$\varphi$: $g(r,\varphi)=\sum_{n=-N}^{+N}g_{n}(r)e^{in\varphi}/2\pi$.
The function $g_{n}(r)$ stands for the $n^{th}$ Fourier coefficient,
and $N+1$ represents the band limitation in the Fourier index $n$: $n<N+1$.
On the one hand, an ideal directional filter has an angular dependence
associated with a $\delta(\varphi)$ distribution, which corresponds
to a null angular width, that is no band limit, $N\rightarrow\infty$,
and constant Fourier coefficients. On the other hand, conceptually,
the basis filters of a steerable filter must have a non-zero angular
width (hence also the filter itself by linearity of the steerability
relation). This ensures that they are sensitive to a whole range of
directions. In that case only, one may imagine to steer the filter
in all directions from a finite number, $M$
in relation (\ref{eq:ds-2}), of these basis filters. The non-zero angular width is naturally
associated with a band limitation $N$ of the steerable filter and
its basis filters. It may be proved rigorously that, if $T$ is the
finite number of non-zero coefficients $g_{n}(r)$ for a filter $g(r,\varphi)$,
then this filter is steerable and $T$ is the minimum number of basis
filters $g_{m}(r,\varphi)$ required in relation (\ref{eq:ds-2})
to steer $g(r,\varphi)$. In other words, the following inequality
holds: $M\geq T$ \cite{STfreeman}. Consequently, if the filter $g(r,\varphi)$
is steerable with a number $M$ of basis filters, then it has a finite
number $T$ of non-zero coefficients and, as suggested here above,
it is inevitably limited in band at some band limit $N$. The notion
of ideal steerable filter ($M$ small, hence $N$ finite) is therefore
clearly in opposition with the notion of ideal directional filter
($N\rightarrow\infty$, hence $M\rightarrow\infty$).

\subsection{Examples of steerable wavelets}

The present subsection introduces the inverse stereographic projection
of derivatives of radial functions on the plane
as examples of steerable wavelets on the sphere.
On the plane, if a sufficiently regular radial function
$\phi(r)$ is considered, its $N^{th}$ derivative in direction $\hat{x}$,
\begin{equation}
\psi^{\partial_{\hat{x}}^{N}}\left(r,\varphi\right)=\partial_{\hat{x}}^{N}\left[\phi\left(r\right)\right],\label{eq:ds-5}
\end{equation}
satisfies the wavelet admissibility condition for any $N\geq1$ and
may be called a wavelet. In terms of the Fourier index $n$ conjugate
to the angular variable $\varphi$, this function is limited in band
at the level $N+1$. It contains $T=N+1$ non-zero coefficients of indices
$n=-N$, $-(N-2)$, ..., $0$, ..., $N-2$, $N$ for $N$ even, and
$n=-N$, $-(N-2)$, ..., $1$, $-1$, ..., $N-2$, $N$ for $N$ odd.
The rotation by an angle $\chi$ of the $N^{th}$ derivative in direction
$\hat{x}$ of a radial function is given as the $N^{th}$ derivative
in direction $\hat{u}(\chi)=r_{\chi}\hat{x}$ : $\psi_{\chi}^{\partial_{\hat{x}}^{N}}(r,\varphi)=\partial_{\hat{u}(\chi)}^{N}[\phi(r)]$.
It may thus be expanded as:
\begin{equation}
\psi_{\chi}^{\partial_{\hat{x}}^{N}}\left(r,\varphi\right)=\sum_{i_{1}...i_{N}=1}^{2}u_{i_{1}}...u_{i_{N}}\psi^{\partial_{x_{i_{1}}}...\partial_{x_{i_{N}}}}\left(r,\varphi\right),
\label{eq:ds-6}
\end{equation}
with $\hat{x}_{1}=\hat{x}$ and $\hat{x}_{2}=\hat{y}$, and where
the coordinates of $\hat{u}(\chi)$ read $(u_{1},u_{2})=(\cos\chi,\sin\chi)$.
It is therefore a steerable filter, expressed in terms of $M=\gamma_{2}^{N}=N+1=T$
combinations with repetitions of the basis filters $\psi^{\partial_{\hat{x}}^{N-k}\partial_{\hat{y}}^{k}}=\partial_{\hat{x}}^{N-k}\partial_{\hat{y}}^{k}[\phi(r)]$,
for $0\leq k\leq N$. In the following paragraphs, we explicitly consider
the examples of the first and second derivatives of a gaussian.

The first derivative of a radial function in direction $\hat{x}$
on the plane reads:
\begin{equation}
\psi^{\partial_{\hat{x}}}\left(r,\varphi\right)=\partial_{r}\phi\left(r\right)\cos\varphi,\label{eq:ds-7}
\end{equation}
where $\partial_{r}$ stands for the radial derivative. It therefore
has an angular band limit at $N+1=2$. It is far from being an ideal
directional filter as its auto-correlation function reads for a normalized
filter: $C^{\partial_{\hat{x}}}(\Delta\chi)=\cos(\Delta\chi)$. In
terms of steerability however, a first derivative is an ideal filter,
as it only requires $N+1=2$ weights. The steerability relation reads,
in terms of the specific rotations $\psi^{\partial_{\hat{x}}}$ and
$\psi^{\partial_{\hat{y}}}$ at $\chi=0$ and $\chi=\pi/2$ respectively:
\begin{equation}
\psi_{\chi}^{\partial_{\hat{x}}}\left(r,\varphi\right)=\psi^{\partial_{\hat{x}}}\left(r,\varphi\right)\cos\chi+\psi^{\partial_{\hat{y}}}\left(r,\varphi\right)\sin\chi.\label{eq:ds-8}
\end{equation}
As a concrete example, let us consider once more the normalized (negative)
first derivative of a gaussian already discussed at the end of § 
\ref{sec:Correspondence-principle}. It is given here through the
general relation (\ref{eq:ds-7}) for $\phi(r)=-\sqrt{2/\pi}e^{-r^{2}/2}$.
The section of this wavelet at $y=0$ is illustrated by Fig. \ref{cap:First-gaussian-steerability-R2-section},
while its section at constant $x$ has a gaussian shape:

\begin{figure}[H]
\begin{center}\includegraphics[width=4cm]{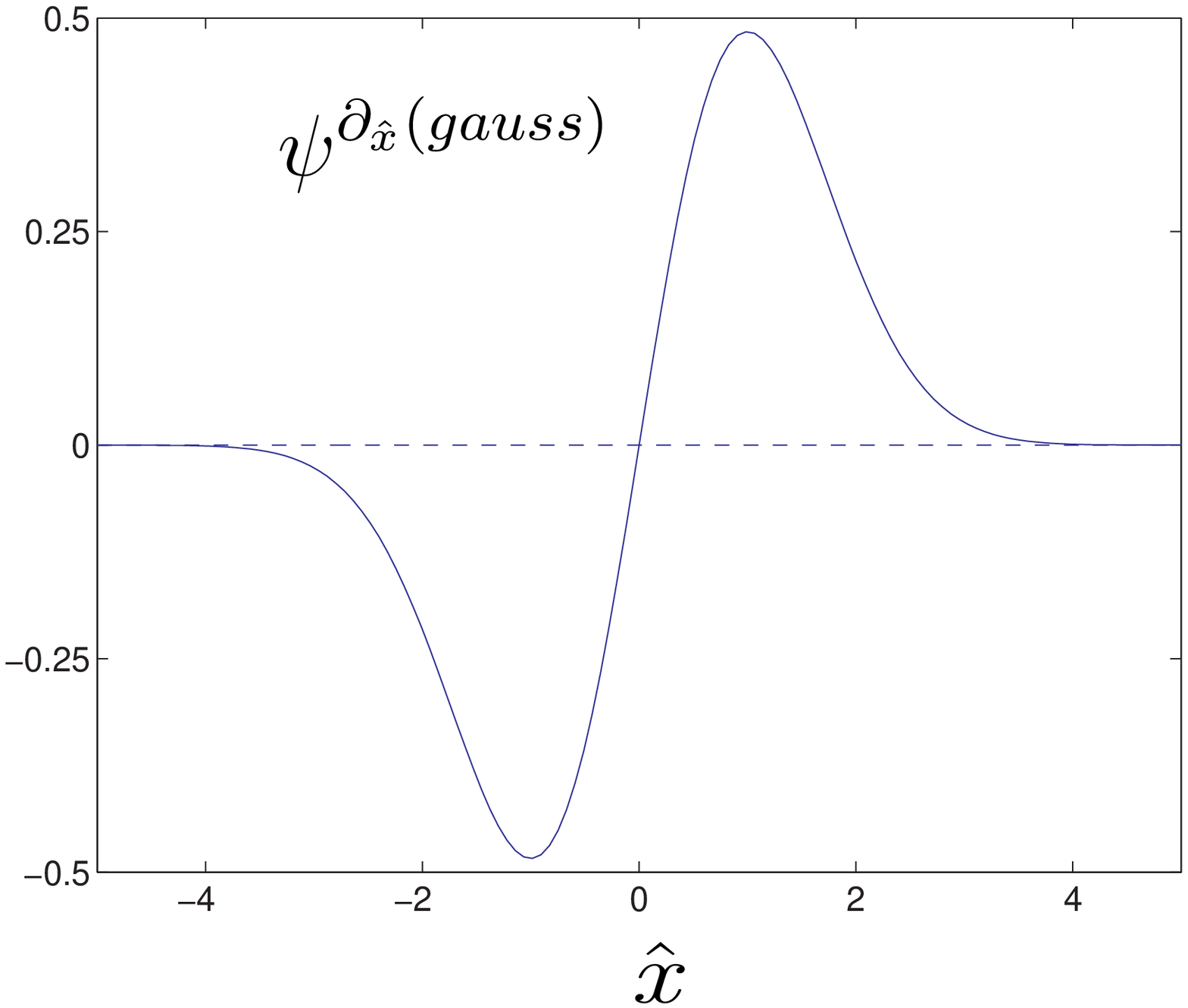}\end{center}

\caption{\label{cap:First-gaussian-steerability-R2-section}Section at $y=0$
of the first derivative of gaussian in direction $\hat{x}$ on the
plane.}
\end{figure}

The inverse stereographic projection on the sphere of the wavelet rotated by $\chi$ and
dilated by $a$ (see relation (\ref{eq:cp-4})), is given, as already discussed, by:
\begin{eqnarray}
\Psi_{\chi,a}^{\partial_{\hat{x}}(gauss)}\left(\theta,\varphi\right) & = & \sqrt{\frac{2}{\pi}}\frac{2}{a^{2}}\left(1+\tan^{2}\frac{\theta}{2}\right)e^{-2\tan^{2}(\theta/2)/a^{2}}\nonumber \\
&  & \left[\tan\frac{\theta}{2}\cos\left(\varphi-\chi\right)\right].\label{eq:ds-9}
\end{eqnarray}
Fig. \ref{cap:First-gaussian-steerability-s2} illustrates the directionality
of first derivatives of radial functions and their steerability relation
(\ref{eq:ds-8}), understood on the sphere, through the specific example of the first gaussian
derivative:

\begin{figure}[H]
\begin{center}\includegraphics[width=2.5cm]{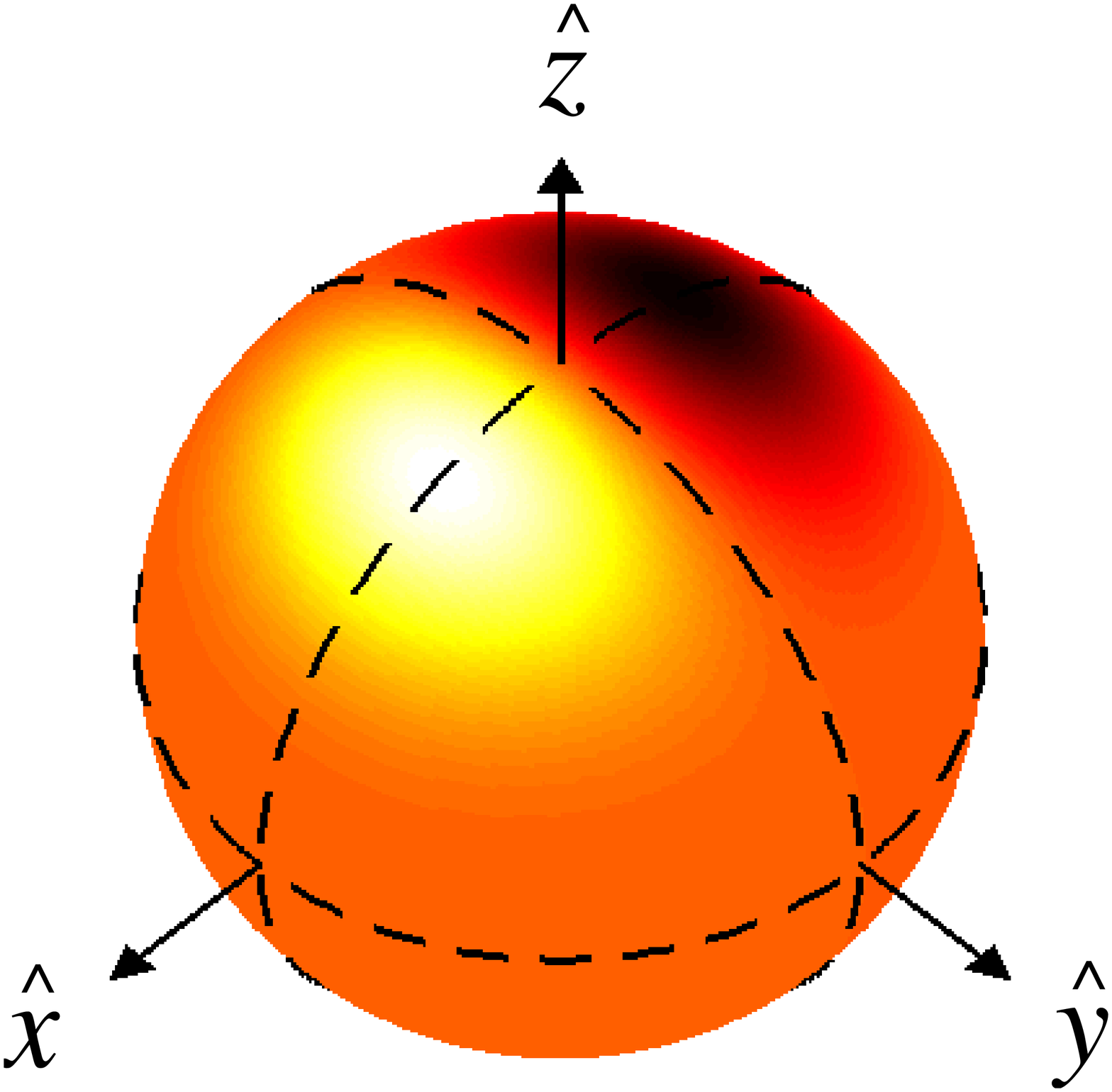}\hspace{2cm}\includegraphics[width=2.5cm]{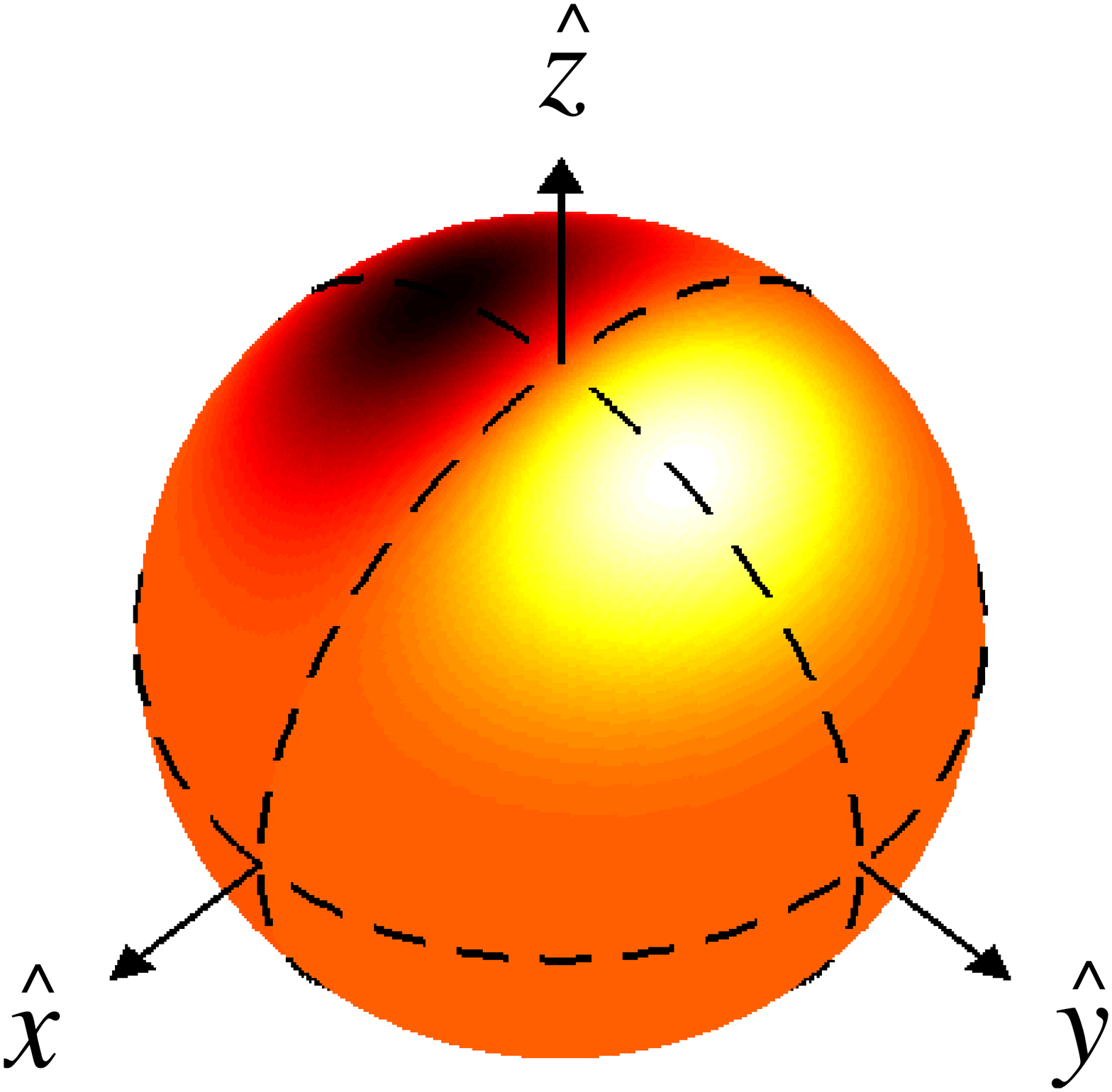}\\
\vspace{0.5cm}\includegraphics[width=2.5cm]{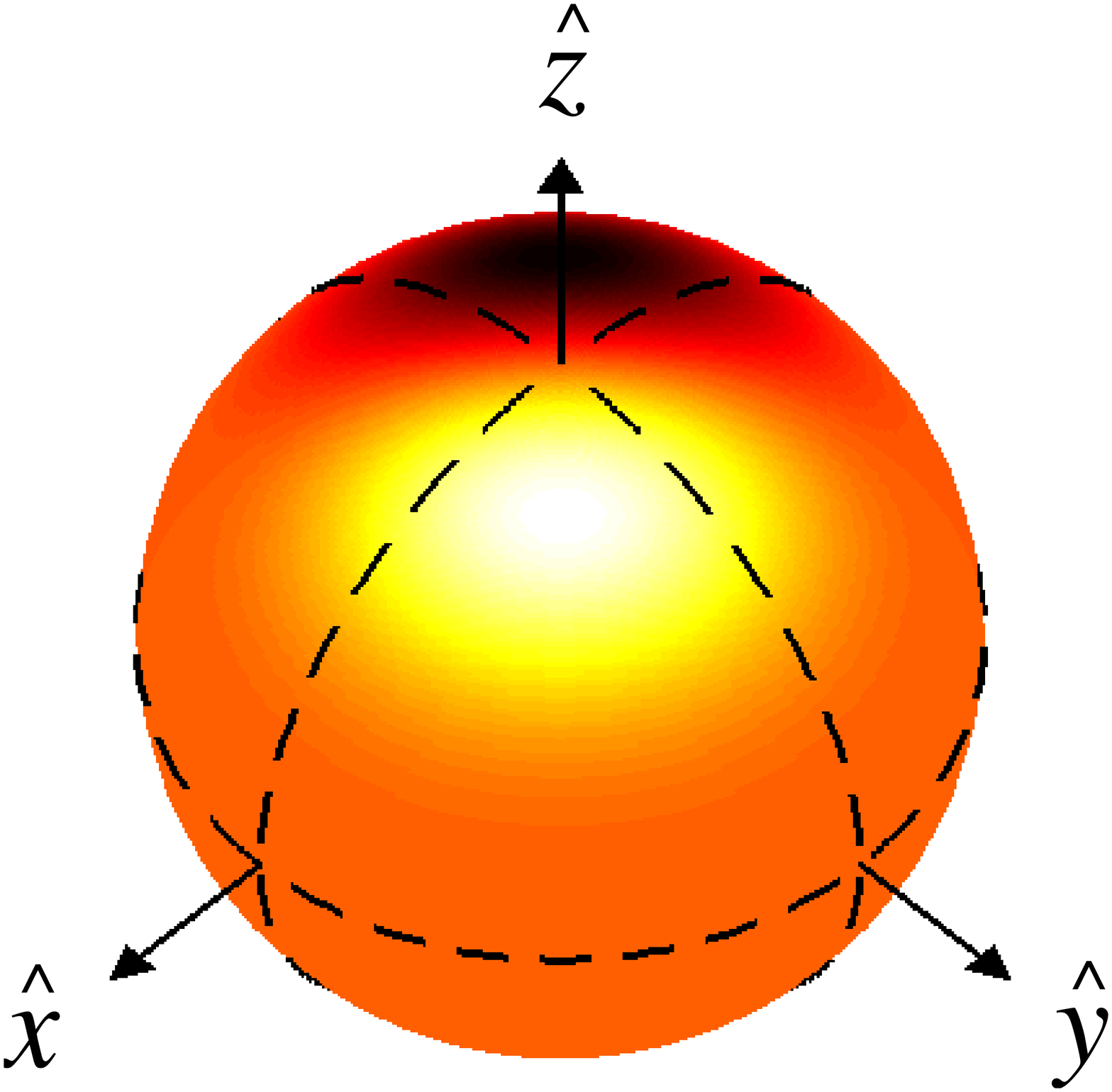}\end{center}

\caption{\label{cap:First-gaussian-steerability-s2}First derivative of gaussian
on the sphere for a dilation factor $a=0.4$: (top left) $\partial_{\hat{x}}$,
(top right) $\partial_{\hat{y}}$, and (bottom) $\partial_{\hat{u}(\chi)}$
with $\chi=\pi/4$. Dark and light regions respectively identify negative
and positive values of the functions.}
\end{figure}

The second derivative of a radial function in direction $\hat{x}$
on the plane reads:
\begin{equation}
\psi^{\partial_{\hat{x}}^{2}}\left(r,\varphi\right)=\frac{1}{r}\partial_{r}\phi\left(r\right)\sin^{2}\varphi+\partial_{r}^{2}\phi\left(r\right)\cos^{2}\varphi.\label{eq:ds-10}
\end{equation}
It is limited in band at $N+1=3$. Its auto-correlation function reads
$C^{\partial_{\hat{x}}^{2}}(\Delta\chi)=A+B\cos2\Delta\chi$, with
the values $A$ and $B$ functions of $\phi(r)$, and $A+B=1$ for
a normalized filter. A second order derivative is therefore not necessarily
a better directional filter than a first order derivative as its auto-correlation
function is not generically better peaked. The general steerability
relation (\ref{eq:ds-2}) holds, the rotated filter being expressed
in terms of basis filters, which are not specific rotations of the
second derivative itself. It requires $N+1=3$ weights:
\begin{eqnarray}
\psi_{\chi}^{\partial_{\hat{x}}^{2}}\left(r,\varphi\right) & = & \psi^{\partial_{\hat{x}}^{2}}\left(r,\varphi\right)\cos^{2}\chi+\psi^{\partial_{\hat{y}}^{2}}
\left(r,\varphi\right)\sin^{2}\chi\nonumber \\
&  & +\psi^{\partial_{\hat{x}}\partial_{\hat{y}}}\left(r,\varphi\right)\sin2\chi.\label{eq:ds-11}
\end{eqnarray}
Again, we consider the example of the normalized (negative) second
derivative of a gaussian, that is for $\phi(r)=-\sqrt{4/3\pi}e^{-r^{2}/2}$
in relation (\ref{eq:ds-10}). The section of this wavelet at $y=0$
is illustrated by Fig. \ref{cap:Second-gaussian-steerability-R2-section},
while its section at constant $x$ has a gaussian shape:

\begin{figure}[H]
\begin{center}\includegraphics[width=4cm]{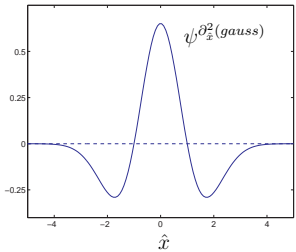}\end{center}

\caption{\label{cap:Second-gaussian-steerability-R2-section}Section at $y=0$
of the second derivative of gaussian in direction $\hat{x}$ on the
plane.}
\end{figure}

The inverse stereographic projection on the sphere of the wavelet
rotated by $\chi$ and dilated by $a$ reads:
\begin{eqnarray}
\Psi_{\chi,a}^{\partial_{\hat{x}}^{2}(gauss)}\left(\theta,\varphi\right) & = & \frac{1}{a}\sqrt{\frac{4}{3\pi}}\left(1+\tan^{2}\frac{\theta}{2}\right)e^{-2\tan^{2}(\theta/2)/a^{2}}\nonumber \\
&  & \left[1-\frac{4}{a^{2}}\tan^{2}\frac{\theta}{2}\cos^{2}\left(\varphi-\chi\right)\right].\label{eq:ds-12}
\end{eqnarray}
Fig. \ref{cap:Second-gaussian-steerability-s2} illustrates the directionality
of second derivatives of radial functions and their steerability relation
(\ref{eq:ds-11}), understood on the sphere, through the specific example of the second gaussian
derivative:

\begin{figure}[H]
\begin{center}\includegraphics[width=2.5cm]{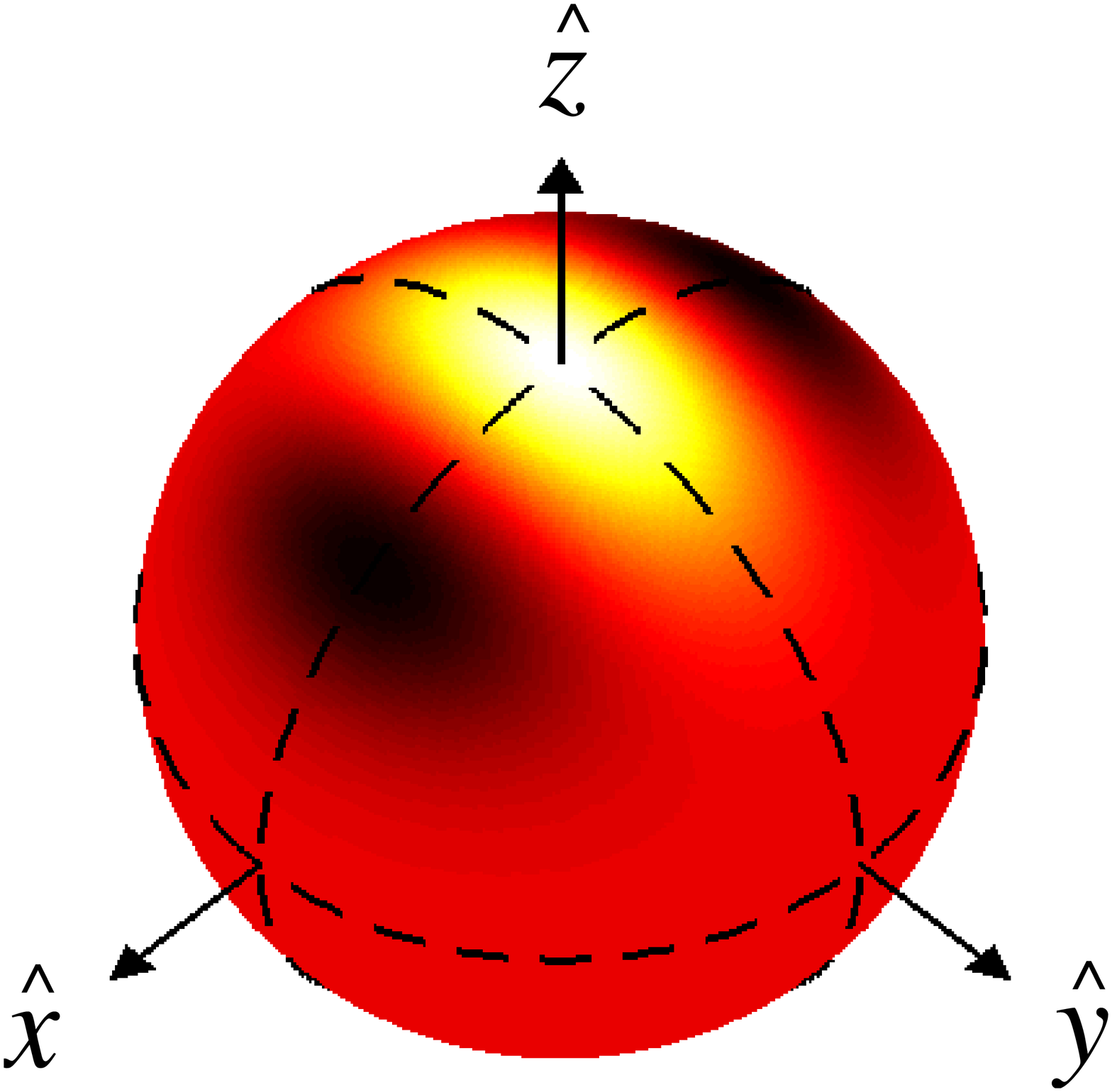}\hspace{0.4cm}\includegraphics[width=2.5cm]{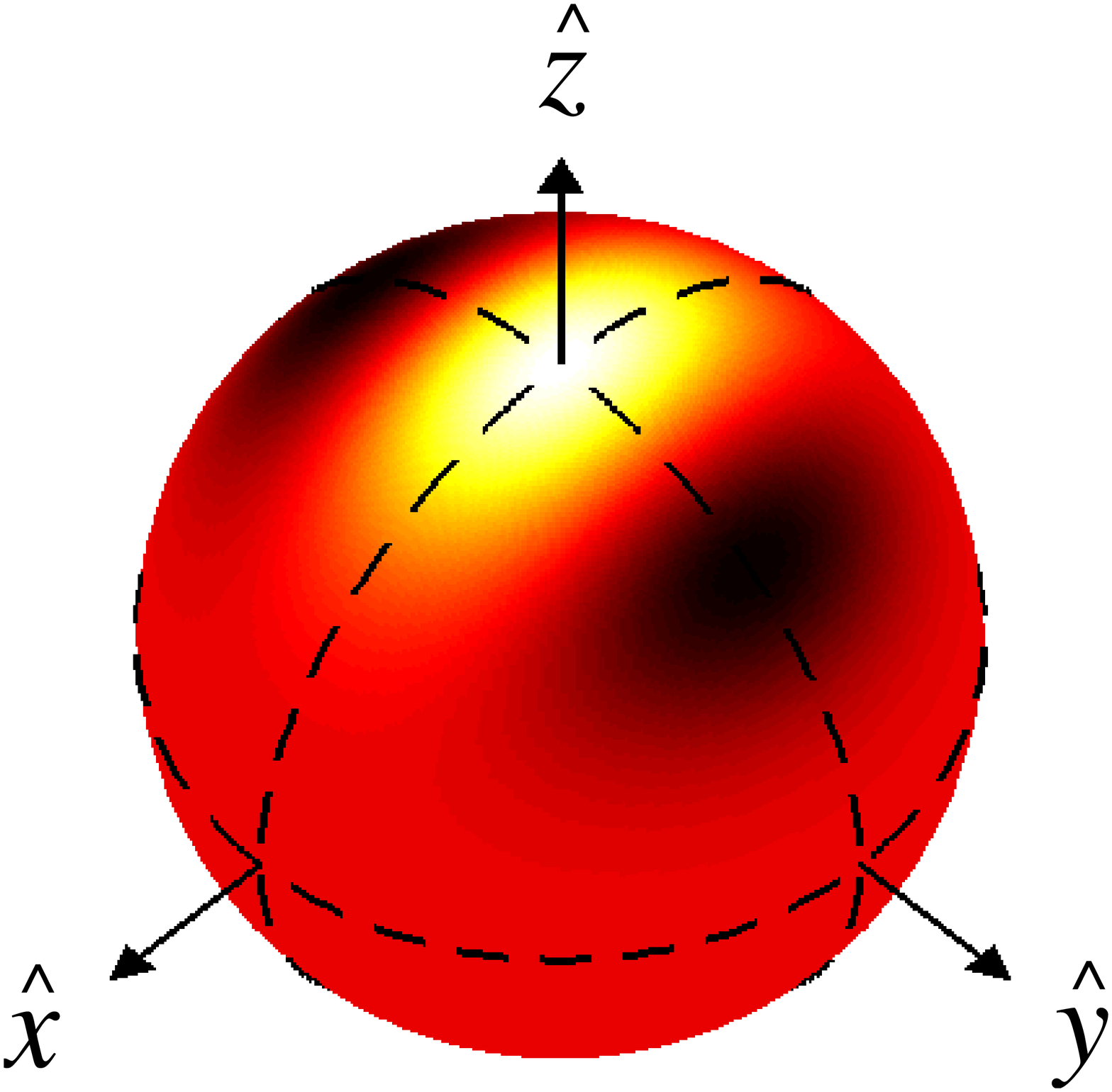}\hspace{0.4cm}\includegraphics[width=2.5cm]{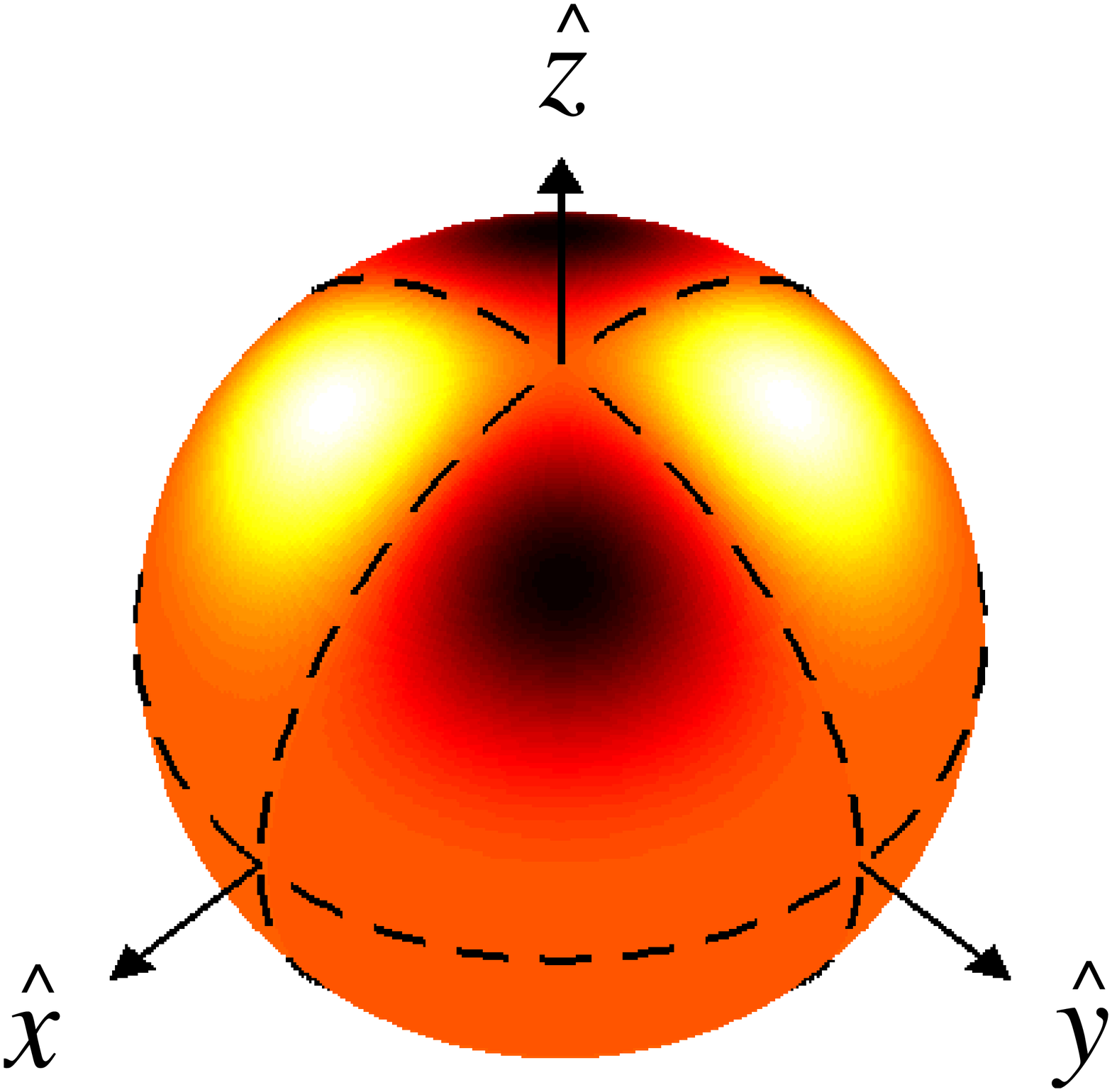}\\
\vspace{0.5cm}\includegraphics[width=2.5cm]{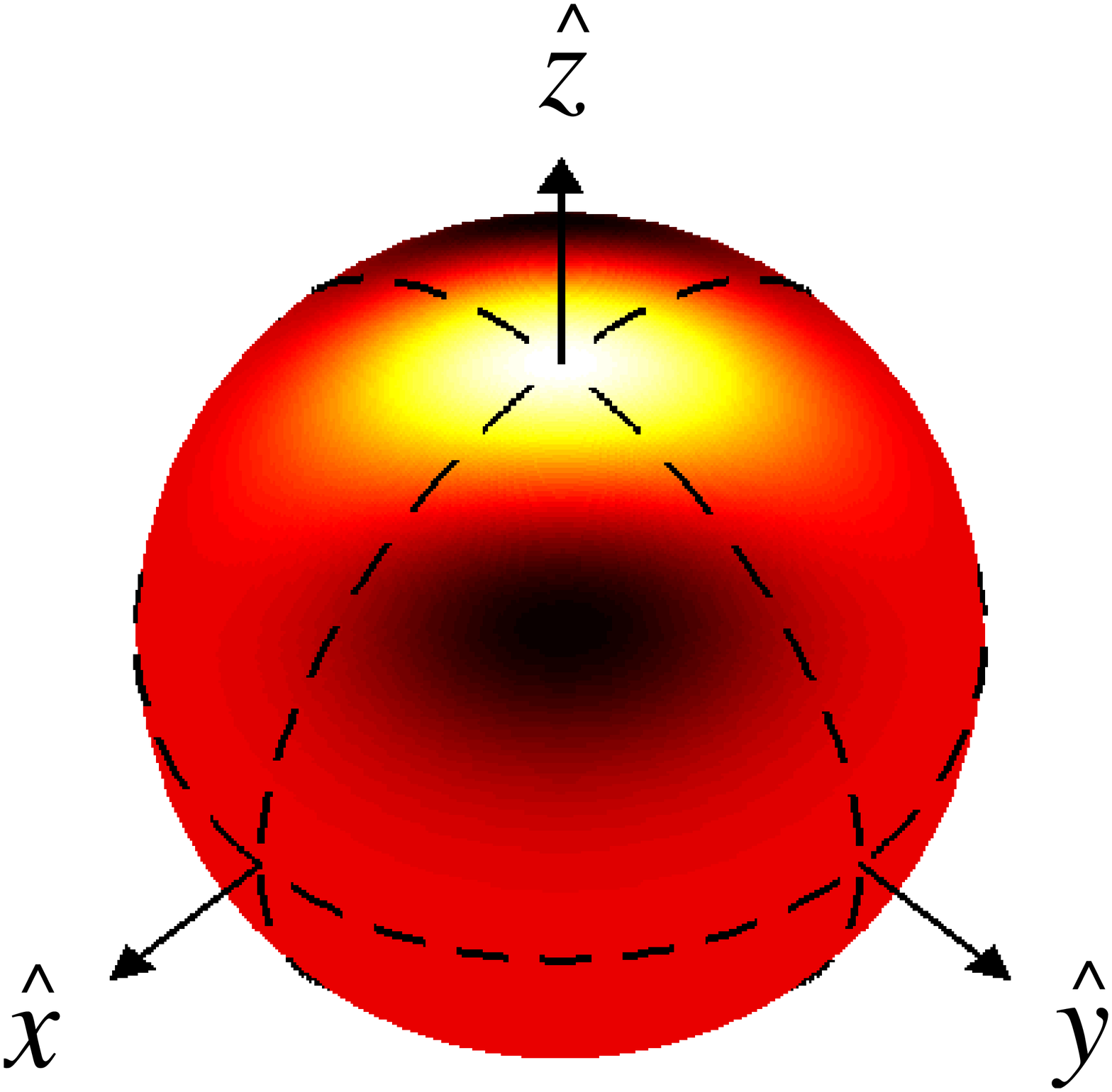}\end{center}

\caption{\label{cap:Second-gaussian-steerability-s2}Second derivative of
gaussian on the sphere for a dilation factor $a=0.4$: (top left)
$\partial_{\hat{x}}^{2}$, (top centre) $\partial_{\hat{y}}^{2}$,
(top right) $\partial_{\hat{x}}\partial_{\hat{y}}$, and (bottom)
$\partial_{\hat{u}(\chi)}^{2}$ with $\chi=\pi/4$. Dark and light
regions again respectively identify negative and positive values of
the functions.}
\end{figure}

\section{CMB local directional features}

\label{sec:CMB-local-directional}In this last section, we discuss
the interest of directional and steerable filters for the identification
and interpretation of local directional features on the sphere, with
direct application to the CMB analysis. First, we show that steerable
filters may efficiently detect local directional features on the sphere
with the same angular precision as ideal directional filters. Second,
the discussion is illustrated by a simple numerical example. Third,
we emphasize on how the property of filter steerability is essential
to reduce the computation cost in the search for directional features
through the wavelet analysis of the CMB.

The theoretical angular resolution power of any directional filter
remains infinite, independently of its possible steerability. This
infinite resolution assumes however that the morphology of the signal
considered is known \emph{a priori}, in particular in the variable
$\varphi$. In the context of the CMB analysis, this might be the
case when looking for the imprint of cosmic strings or other pre-defined
features in the background radiation. In that case indeed, the wavelet
coefficient is known analytically as a pure function of the difference
$\Delta\chi=\chi-\chi^{*}$ between the wavelet rotated by $\chi$,
and the direction $\chi^{*}$ of the considered feature. This direction
$\chi^{*}$ may therefore be identified exactly in the analysis of
the wavelet coefficients of the signal. Notice however that for a
given experimental precision, the resolution power of filters is directly
related to their directionality, through the peakedness of their auto-correlation
function, and is therefore a function of the angular band limitation
and the number of weights $M$ in the case of steerable filters.

The following simple example illustrates the theoretical infinite
angular resolution power of steerable filters.

\begin{figure}[H]
\begin{center}\includegraphics[width=4.5cm]{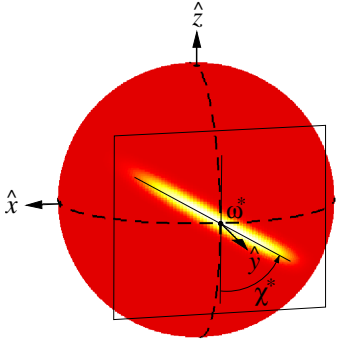}\end{center}

\caption{\label{cap:Signal-example}Test-signal: elongated feature centered
at the point $\omega^{*}=(\theta^{*},\varphi^{*})=(\pi/2,\pi/2)$
of the sphere, and making an angle $\chi^{*}=\pi/3$ with the meridian
at that point.}
\end{figure}

\begin{figure}[H]
\begin{center}\includegraphics[width=4.5cm]{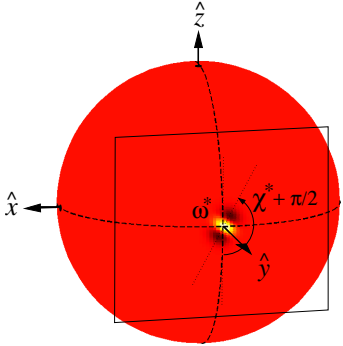}\end{center}

\caption{\label{cap:Selectedwavelet-example}Analysis-wavelet: second gaussian
derivative in direction $\hat{x}$, projected on the sphere. It is shown centered
at the point $\omega^{*}=(\theta^{*},\varphi^{*})=(\pi/2,\pi/2)$
of the sphere, at a scale $a=0.05$, and making an angle $\chi_{0}=\chi^{*}+\pi/2=5\pi/6$
with the meridian at that point, that is in the configuration which
maximizes the corresponding wavelet coefficient of the test-signal.}
\end{figure}

First, the test-signal is defined as an elongated feature centered
at the point $\omega^{*}=(\theta^{*},\varphi^{*})=(\pi/2,\pi/2)$
of the sphere, and making an angle $\chi^{*}=\pi/3$ with the meridian
at that point. The signal is analytically defined
by the function $F(\theta,\varphi)=\exp[-(\cos\chi^{*}x-\sin\chi^{*}z)^{2}/2\sigma^{2}]\exp[-16(y-1)^{2}]$,
with the Cartesian coordinates of points of the sphere related to
their spherical coordinates by $(x,y,z)=(\sin\theta\cos\varphi,\sin\theta\sin\varphi,\cos\theta)$,
and for a half width $\sigma=0.05$ (see Fig. \ref{cap:Signal-example}).
Second, the signal is analyzed
by its correlation with a second derivative of a gaussian $\Psi^{\partial_{\hat{x}}^{2}}$,
that is in terms of its wavelet coefficients $W_{^{\partial_{\hat{x}}^{2}}}^{F}(\omega_{0},\chi,a)$
at each point $\omega_{0}$ of the sphere, and for any scale $a$
and direction $\chi$ of the wavelet. We might have equivalently chosen
the first gaussian derivative or any other steerable wavelet. Notice
that the zeros of the second gaussian derivative at the North pole
are located at $\theta_{0}=2\arctan(a/2)$ in the direction of the
wavelet $\varphi=\chi$, and in the opposite direction $\varphi=\chi+\pi$
(see equation (\ref{eq:ds-12}), and Fig. (\ref{cap:Second-gaussian-steerability-s2})).
This angular opening $\theta_{0}$ may be understood as a qualitative
measure of the half width of the wavelet. For a dilation factor $a=1$,
we get $\theta_{0}\simeq0.9\,\textnormal{rad}$. For a small dilation
factor $a$, the half width is simply given as $\theta_{0}\simeq a\,\textnormal{rad}$.
We consider a highly localized analysis-wavelet for a dilation factor
$a=0.05$, which corresponds to a typical half width $\theta_{0}\simeq0.05\,\textnormal{rad}$.
It is therefore of the same size as the width of the feature in the
direction $\chi^{*}+\pi/2$ as $a=\sigma=0.05$, but much smaller
than its elongation in the direction $\chi^{*}$. At the North pole,
the non-rotated second gaussian derivative with $a=0.05$ has essentially
zero mean in direction $\hat{x}$ (this is only exact in the Euclidean
limit, that is $a\rightarrow0$), and a maximum mean in direction
$\hat{y}$. At any point $\omega_{0}$ on the signal, the wavelet
transform is therefore minimum if the wavelet is directed along the
signal, which is essentially constant on an interval of the size of
the wavelet. On the contrary, it is maximum if it is directed at an
angle $\pi/2$ relative to the feature, direction in which it has
the same scale as the signal itself (see Fig. \ref{cap:Selectedwavelet-example}).

\begin{figure}[H]
\begin{center}\includegraphics[width=6cm]{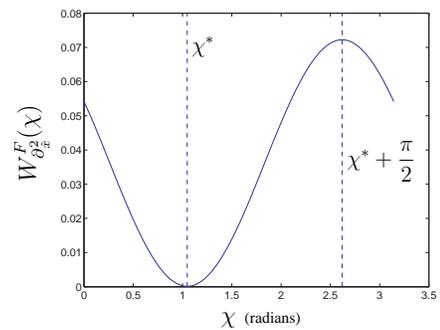}\end{center}

\caption{\label{cap:The-wavelet-coefficient}Wavelet coefficient as a function
of $\chi$ for the considered elongated feature, analyzed with a second
gaussian derivative at the point $\omega_{0}=(\pi/2,\pi/2)$ and at
the chosen scale $a=0.05$.}
\end{figure}

Through the correspondence principle and the linearity of the wavelet
filtering, the steerability relation (\ref{eq:ds-11}) for the second
derivative of a gaussian may be equivalently written for the spherical
wavelet coefficients as:
\begin{equation}
W_{_{\partial_{\hat{x}}^{2}}}^{F}\left(\chi\right)=W_{\partial_{\hat{x}}^{2}}^{F}\cos^{2}\chi+W_{^{\partial_{\hat{y}}^{2}}}^{F}\sin^{2}\chi+
W_{_{\partial_{\hat{x}}\partial_{\hat{y}}}}^{F}\sin2\chi,\label{eq:cmb-1}
\end{equation}
where we dropped the dependence of each coefficient in $\omega_{0}$
and $a$. Each basis coefficient is evaluated at $\chi=0$ for the
corresponding basis wavelet. This relation may also be written in
a form similar to the expression of the auto-correlation function
of the second gaussian derivative. This is natural as the auto-correlation
function is nothing else but the particular wavelet coefficient of
the rotated wavelet analyzed by itself. We get indeed $W_{^{\partial_{\hat{x}}^{2}}}^{F}(\chi)=A'+B'\cos2(\chi-\chi_{0})$,
for $A'=(W_{\partial_{\hat{x}}^{2}}^{F}+W_{\partial_{\hat{y}}^{2}}^{F})/2$
and $B'=(W_{\partial_{\hat{x}}^{2}}^{F}-W_{\partial_{\hat{y}}^{2}}^{F})/2\cos\chi_{0}$,
with a maximum at $\chi_{0}$ defined by
\begin{equation}
\tan2\chi_{0}=\frac{2W_{\partial_{\hat{x}}\partial_{\hat{y}}}^{F}}{W_{\partial_{\hat{x}}^{2}}^{F}-W_{\partial_{\hat{y}}^{2}}^{F}}.\label{eq:cmb-2}
\end{equation}
At each point $\omega_{0}$ considered, the cost of the analysis
is therefore reduced to the computation of the three wavelet coefficients
for the basis wavelets at the chosen scale $a=0.05$, and in their
original direction $\chi=0$. The direction $\chi^{*}$ of the feature
at that point is given as $\chi^{*}=\chi_{0}-\pi/2$. Fig. \ref{cap:The-wavelet-coefficient}
represents the wavelet coefficient $W_{_{\partial_{\hat{x}}^{2}}}^{F}(\chi)$
as a function of $\chi$, resulting from this directional analysis
at the point $\omega_{0}=\omega^{*}=(\pi/2,\pi/2)$, through the relation
(\ref{eq:cmb-1}).

The direction of the feature is precisely recovered at $\chi^{*}=\pi/3$,
up to negligible numerical errors. This clearly illustrates the fact
that steerable wavelets have an infinite angular resolution power
for an ideal experiment where no error affects the signal.

In conclusion, the steerability of wavelets is a property of fundamental
interest in the analysis of local directional features on the sphere.
First, the filter steerability allows a drastic reduction in computation
cost for analyzing all possible directions of features at each point
on the sphere, as only a small number $M$ of basis directions must
be considered at each point. Second, as the above example illustrates,
this reduction is achieved theoretically without any precision loss
in the identification of the directions. In the context of the CMB
study, let us recall that local directional features may be associated
with non-gaussianity, statistical anisotropy, or foreground emission.
Their identification requires for each analysis scale $a$, the correlation
of a filter in all directions $\chi$ and at all points $\omega_{0}$
of the sphere with numerous theoretical simulations of the signal.
Such an analysis is currently hardly affordable in terms of computation
time at the already high resolution level of the CMB maps, when one
wants the same precision in the identification of the direction ($>10^{3}$
sampling points) as of the localization ($10^{6}$ sampling points
on the sphere) of singular features \cite{CVmcewen,WNGmcewen}. The
property of filter steerability may therefore be used to reduce the
complexity of such calculations and eventually render them accessible.
However, Fig. \ref{cap:The-wavelet-coefficient} also illustrates
the non optimal peakedness of the wavelet coefficients of the considered
signal as a function of directions ($W_{^{\partial_{\hat{x}}^{2}}}^{F}(\chi)=A'+B'\cos2(\chi-\chi_{0})$),
equivalently to the peakedness of the wavelet auto-correlation. As
already discussed, in a practical situation, a better directionality
of the filter would enhance the stability of the measurement of directions
relatively to noise. The specific choice of filter, optimizing the
compromise between ideal directionality and steerability, will depend
very much on the application considered, in terms of noise, precision
requirements, and computational resources.

\section{Conclusion}

\label{sec:Conclusion}The recent developments in the analysis of
the cosmic microwave background (CMB) radiation ask for new methods
of scale-space signal analysis on the sphere, notably for the identification
of foreground emission, or for testing the hypotheses of gaussianity
and statistical isotropy of the CMB, on which the concordance cosmological
model relies today.

In the work presented here, we reintroduced the formalism of wavelets
on the sphere in a practical and self-consistent approach, simply
understanding wavelets as localized filters which enable a scale-space
analysis and provide an explicit reconstruction of the signal considered
from its wavelet coefficients. We also proved a correspondence principle
which states that the inverse stereographic projection of a wavelet
on the plane gives a wavelet on the sphere, and allows to transfer
wavelet properties from the plane onto the sphere. In that context,
we finally defined and discussed the notions of directionality and
steerability of filters on the sphere for the analysis of local directional
features in the signal considered.

The practical formalism introduced provides a method for the detection
of local features in the CMB, and the identification of their precise
direction, through the analysis of the wavelet coefficients of the
signal. Notice however that these generic developments of signal processing
on the sphere may find numerous applications beyond cosmology and
astrophysics, as soon as the data to be analyzed are distributed on
the sphere.

\acknowledgements

The authors wish to thank J.-P. Antoine and P. Vielva for valuable
comments and discussions. They thank the referee for very constructive
remarks. They acknowledge support of the HASSIP (Harmonic Analysis and
Statistics for Signal and Image Processing) European research network
(HPRN-CT-2002-00285). The work of Y. W. is also supported by the Swiss
National Science Foundation.

\appendix

\section{Technical proofs for wavelets on the sphere}

\label{sec:Wavelets-sphere-app}In this first appendix, we establish
technical proofs for the wavelet formalism on the sphere defined in §
\ref{sec:Wavelets-sphere}. First, we
prove the uniqueness of the unitary, radial, and conformal operator
$D(a)$ of dilation on functions in $L^{2}(S^{2},d\Omega)$ on the
sphere, and give its expression. Second, we explicitly prove the admissibility
condition for wavelets on the sphere (\ref{eq:ws-6}). This condition
is required to ensure the exact reconstruction formula (\ref{eq:ws-5})
for any signal $F(\omega)$ from its wavelet coefficients (\ref{eq:ws-4}),
in a decomposition with the wavelet $\Psi(\omega)$ in $L^{2}(S^{2},d\Omega)$.

\subsection{Uniqueness of the dilation operator}

The dilation operator $D(a)$ on functions in $L^{2}(S^{2},d\Omega)$
is defined in terms of the inverse of the corresponding dilation $D_{a}$
on points in $S^{2}$, applied to the argument of the function considered.
The dilation operator $D_{a}$ must be a radial and conformal diffeomorphism.
A radial operator only affects the radial variable $\theta$ independently
of $\varphi$, and leaves $\varphi$ invariant. This property is natural
to define a dilation of points on the sphere relative to the North
pole. In this context, the operator $D(a)$ on functions in $L^{2}(S^{2},d\Omega)$
takes a generic form given by relation (\ref{eq:ws-3}), in terms
of a radial diffeomorphism $D_{a}(\theta,\varphi)=(\theta_{a}(\theta),\varphi)$
on $S^{2}$. The diffeomorphism $\theta_{a}(\theta)$ and the function
$\lambda^{1/2}(a,\theta)$ have to be determined from the additional
properties required.

First, the diffeomorphism $D_{a}$ may be understood as a coordinate
transformation on the sphere, $D_{a}:\omega=(x^1,x^2)=(\theta,\varphi)\rightarrow
\omega'=(x'^1,x'^2)=(\theta_{a},\varphi)\in S^2$.
It must be conformal in order to preserve the measure of angles and directions,
which are defined locally in the tangent plane at each point of $S^2$.
This explicitly requires that the transformed metric
$g'_{ij}(\theta_{a},\varphi)=(\partial x^k/\partial x'^i)
(\partial x^l/\partial x'^j)g_{kl}(\theta(\theta_{a}),\varphi)$ is conformally equivalent to the original
metric $g_{ij}(\theta_{a},\varphi)$ on the sphere, with $i,j,k,l\in\{1,2\}$.
We consider the metric induced on the sphere from the Euclidean metric
in three dimensions: $g_{ij}(\theta,\varphi)=\textnormal{diag}(1,\sin^{2}\theta)$.
The conformal equivalence reads by definition: $g'_{ij}(\theta_{a},\varphi)=e^{\phi(a,\theta(\theta_{a}))}g_{ij}(\theta_{a},\varphi)$,
for some strictly positive conformal factor $e^{\phi(a,\theta)}$,
that is for a real function $\phi(a,\theta)$. This condition implies
that the operator $D_{a}$ is linear in $\tan\theta/2$: $\tan\theta_{a}(\theta)/2=\alpha(a)\tan\theta/2$.
The function $\alpha(a)$ must be strictly increasing in the dilation
factor $a\in\mathbb{R}_{+}^{*}$, with the limits $\alpha(0)=0$ and
$\alpha(\infty)=\infty$. The group structure for the composition
of dilations must also be fulfilled. This constrains $\alpha(a)$
to $\alpha(a)=a^{\alpha_{0}}$ for a strictly positive exponent
$\alpha_{0}\in\mathbb{R}_{+}^{*}$.
We take, without loss of generality, a function linear in the
dilation factor $a$, corresponding to $\alpha_{0}=1$:
\begin{equation}
\tan\frac{\theta_{a}\left(\theta\right)}{2}=a\tan\frac{\theta}{2}.\label{eq:a-ws-1}
\end{equation}
Any other choice for $\alpha_{0}$ would simply correspond to a rescaling
of the dilation factor. By dilation, the sphere without its South pole is thus
mapped on itself: $\theta_a(\theta):\theta\in [0,\pi[\rightarrow\theta_a\in [0,\pi[$.
The conformal equivalence also determines
the conformal factor: $e^{-\phi(a,\theta)/2}=a[1+\tan^{2}(\theta/2)]/[(1+a^{2}\tan^{2}(\theta/2))]$.

Second, the unitarity of the dilation operator $D(a)$ on functions
in $L^{2}(S^{2},d\Omega)$ identifies $\lambda(a,\theta)$ with the
conformal factor: $\lambda(a,\theta)=e^{\phi(a,\theta_{1/a}(\theta))}=e^{-\phi(a^{-1},\theta)}$,
or equivalently $\lambda^{1/2}(a,\theta)=a^{-1}[1+\tan^{2}(\theta/2)]/[(1+a^{-2}\tan^{2}(\theta/2))]$.
This identity simply relies on the relation between the invariant
measure on the sphere and the measure obtained after the conformal
transformation $D_{a}$: $\sin\theta_{a}(\theta)d\theta_{a}(\theta)d\varphi=e^{-\phi(a,\theta)}\sin\theta d\theta d\varphi$.

On the plane, the construction of the dilation $d(a)$ on functions,
unique unitary, radial, and conformal operator in $L^{2}(\mathbb{R}^{2},d^{2}\vec{x})$,
relies on an identical reasoning. The canonical Euclidean metric on the plane in
polar coordinates simply reads $g_{ij}(\theta,\varphi)=\textnormal{diag}(1,r^{2})$,
for $i,j\in\{ r,\varphi\}$. The expression (\ref{eq:wp-3}) trivially
follows.

\subsection{Establishment of the admissibility condition}

First, we recall the expressions for the Fourier decomposition of
a function $G(\omega)$ in $L^{2}(S^{2},d\Omega)$ on the sphere.
The spherical harmonics $Y_{lm}(\omega)$, with $l\in\mathbb{N}$,
$m\in\mathbb{Z}$, and $|m|\leq l$, form an orthonormal basis
in $L^{2}(S^{2},d\Omega)$:
\begin{equation}
\int_{S^{2}}d\Omega\, Y_{lm}^{*}\left(\omega\right)Y_{l'm'}\left(\omega\right)=\delta_{ll'}\delta_{mm'}.\label{eq:a-ws-2}
\end{equation}
The direct and inverse spherical harmonics transforms of a function
$G(\omega)$ are respectively defined as:\begin{eqnarray}
\widehat{G}_{lm} & = & \int_{S^{2}}d\Omega\, Y_{lm}^{*}\left(\omega\right)G\left(\omega\right)\label{eq:a-ws-3}\\
G\left(\omega\right) & = & \sum_{l\in\mathbb{N}}\sum_{|m|\leq l}\widehat{G}_{lm}Y_{lm}\left(\omega\right).\label{eq:a-ws-4}\end{eqnarray}

Second, we introduce the transformation law of functions on the sphere
under rotation, in terms of the Wigner $D$-functions \cite{CVbrink}.
Let $\rho$ be an element of $SO(3)$, with $\rho=(\varphi,\theta,\chi)$,
in a decomposition in the Euler angles $\varphi$, $\theta$, and
$\chi$. The Wigner $D$-functions $D_{mn}^{l}(\rho)$, with $l\in\mathbb{N}$,
$m,n\in\mathbb{Z}$, and $|m|,|n|\leq l$, are the matrix elements
of the irreducible unitary representations of weight $l$ of the rotation
group in the space of square-integrable functions $L^{2}(SO(3),d\rho)$
on $SO(3)$, with the invariant measure $d\rho=d\varphi d\cos\theta d\chi$.
By the Peter-Weyl theorem on compact groups, the matrix elements $D_{mn}^{l*}$
also form an orthogonal basis in $L^{2}(SO(3),d\rho)$,
with the following orthogonality relation:
\begin{equation}
\int_{SO(3)}d\rho\, D_{mn}^{l}\left(\rho\right)D_{m'n'}^{l'*}\left(\rho\right)=\frac{8\pi^{2}}{2l+1}\delta_{ll'}\delta_{mm'}\delta_{nn'}.\label{eq:a-ws-5}
\end{equation}
Let us consider the decomposition $\rho=(\omega_{0},\chi)$, with
$\omega_{0}=(\theta_{0},\varphi_{0})$ identifying a point on the
sphere $S^{2}$, and $\chi\in[0,2\pi[$ identifying a direction at
each point. The action of the corresponding operator $R(\omega_{0},\chi)$
on a function $G(\omega)$ in $L^{2}(S^{2},d\Omega)$ on the sphere
reads in terms of its spherical harmonics coefficients and the Wigner
$D$-functions:
\begin{equation}
\widehat{\left[R\left(\omega_{0},\chi\right)G\right]}_{lm}=\sum_{|n|\leq l}D_{mn}^{l}\left(\omega_{0},\chi\right)\widehat{G}_{ln}.\label{eq:a-ws-6}
\end{equation}

Finally, we prove the admissibility condition for a wavelet on the
sphere. Let $F(\omega)$ be a function in $L^{2}(S^{2},d\Omega)$
on the sphere, and $\Psi(\omega)$ in $L^{2}(S^{2},d\Omega)$, the
wavelet considered for the decomposition into the coefficients $W_{\Psi}^{F}(\omega_{0},\chi,a)$
given in (\ref{eq:ws-4}). We want to establish the condition under
which the explicit reconstruction formula (\ref{eq:ws-5}) holds.
From the expressions (\ref{eq:a-ws-4}) and (\ref{eq:a-ws-6}) and
the definition of the operator $L_{\Psi}$, the function $[R(\omega_{0},\chi)L_{\Psi}\Psi_{a}](\omega)=[L_{\Psi}\Psi_{a}](R_{\omega_{0},\chi}^{-1}\omega)$
takes the form
\begin{equation}
\left[R\left(\omega_{0},\chi\right)L_{\Psi}\Psi_{a}\right]\left(\omega\right)=\sum_{l\in\mathbb{N}}\sum_{|m|,|n|\leq l}\frac{1}{C_{\Psi}^{l}}D_{mn}^{l}\left(\omega_{0},
\chi\right)\widehat{\left(\Psi_{a}\right)}_{ln}Y_{lm}\left(\omega\right).\label{eq:a-ws-7}
\end{equation}
The wavelet coefficient $W_{\Psi}^{F}(\omega_{0},\chi,a)$ defined
in (\ref{eq:ws-4}) may be written as:
\begin{equation}
W_{\Psi}^{F}\left(\omega_{0},\chi,a\right)=\sum_{l\in\mathbb{N}}\sum_{|m|,|n|\leq l}D_{mn}^{l*}\left(\omega_{0},\chi\right)\widehat{\left(\Psi_{a}\right)}_{ln}^{*}\widehat{F}_{lm}.
\label{eq:a-ws-8}
\end{equation}
Inserting these last two expressions in (\ref{eq:ws-5}), and using
the orthogonality relation (\ref{eq:a-ws-5}) for the Wigner $D$-functions,
we obtain:
\begin{equation}
F\left(\omega\right)=\sum_{l\in\mathbb{N}}\sum_{|m|\leq l}\widehat{F}_{lm}Y_{lm}\left(\omega\right)\frac{1}{C_{\Psi}^{l}}\left[\frac{8\pi^2}{2l+1}
\sum_{|n|\leq l}\int_{0}^{+\infty}\frac{da}{a^{3}}\,|\widehat{\left(\Psi_{a}\right)}_{ln}|^{2}\right].\label{eq:a-ws-9}
\end{equation}
From this last expression it is obvious that the reconstruction formula
(\ref{eq:ws-5}) holds if and only if the coefficients $C_{\Psi}^{l}$
defined in (\ref{eq:ws-6}) are finite and non-zero for any $l\in\mathbb{N}$.
This explicitly establishes the wavelet admissibility condition (\ref{eq:ws-6})
on the sphere.

\section{Technical proofs for the correspondence principle}

\label{sec:Correspondence-principle-app}In this second appendix,
we prove the correspondence principle defined in §
\ref{sec:Correspondence-principle},
which states that the inverse stereographic projection of a wavelet
$\psi(\vec{x})$ in $L^{2}(\mathbb{R}^{2},d^{2}\vec{x})$ on the plane,
thus satisfying the admissibility condition (\ref{eq:wp-6}), gives
a wavelet $\Psi(\omega)$ in $L^{2}(S^{2},d\Omega)$ on the sphere,
satisfying the admissibility condition (\ref{eq:ws-6}). First, we
reformulate the admissibility conditions on the plane and on the sphere.
Second, we establish explicitly that the projection of a wavelet on
the plane gives a wavelet on the sphere, in terms of these reformulated
admissibility conditions. In that respect, we only require that the
projection $\Pi$ between functions in $L^{2}(S^{2},d\Omega)$ and
in $L^{2}(\mathbb{R}^{2},d^{2}\vec{x})$ be a unitary, radial, and
conformal diffeomorphism. Finally, we prove that the stereographic
projection is the unique projection operator satisfying these properties.

\subsection{Reformulation of the admissibility conditions}

First, let $\psi(\vec{x})$ be a function in $L^{2}(\mathbb{R}^{2},d^{2}\vec{x})$.
We show that the wavelet admissibility condition (\ref{eq:wp-6})
on the plane applied to $\psi(\vec{x})$ is equivalent to the condition:
\begin{equation}
0<I_{\psi}^{f}=\int_{0}^{2\pi}d\chi\int_{0}^{+\infty}\frac{da}{a^{3}}\int_{\mathbb{R}^{2}}d^{2}\vec{x}_{0}\,|\langle\psi_{\vec{x}_{0},\chi,a}|f\rangle|^{2}<\infty,\label{eq:a-cp-1}
\end{equation}
for any $f(\vec{x})\neq0$ in $L^{2}(\mathbb{R}^{2},d^{2}\vec{x})$.
To prove this statement, we simply notice that the scalar product
is preserved up to a factor $2\pi$ by the Fourier transform. This
implies that $2\pi\langle\psi_{\vec{x}_{0},\chi,a}|f\rangle=\langle\widehat{(\psi_{\vec{x}_{0},\chi,a})}|\hat{f}\rangle$,
with $\widehat{(\psi_{\vec{x}_{0},\chi,a})}(\vec{k})=a\hat{\psi}(ar_{\chi}^{-1}\vec{k})e^{-i\vec{k}\cdot\vec{x}_{0}}$.
Using the orthogonality relation of the imaginary exponentials, the
integral $I_{\psi}^{f}$ therefore takes the form
\begin{equation}
I_{\psi}^{f}=\left[\int_{0}^{2\pi}d\chi\int_{0}^{+\infty}\frac{da}{a}\,|\widehat{\Psi}\big(ar_{\chi}^{-1}\vec{k}\big)|^{2}\right]||f||^{2},\label{eq:a-cp-2}
\end{equation}
where $||f||$ stands for the norm of $f(\vec{x})$ in $L^{2}(\mathbb{R}^{2},d^{2}\vec{x})$.
A simple change of variable in the integrals leads to the equality
\begin{equation}
I_{\psi}^{f}=C_{\psi}||f||^{2},\label{eq:a-cp-3}
\end{equation}
which proves the equivalence between the condition (\ref{eq:a-cp-1})
and the wavelet admissibility condition (\ref{eq:wp-6}) on the plane.

Second, let $\Psi(\omega)$ be a function in $L^{2}(S^{2},d\Omega)$.
We show in a similar way that the wavelet admissibility condition
(\ref{eq:ws-6}) on the sphere applied to $\Psi(\omega)$ is induced by the condition:
\begin{equation}
0<I_{\Psi}^{F}=\int_{0}^{2\pi}d\chi\int_{0}^{+\infty}\frac{da}{a^{3}}\int_{S^{2}}d\Omega_{0}\,|\langle\Psi_{\omega_{0},\chi,a}|F\rangle|^{2}<\infty,\label{eq:a-cp-4}
\end{equation}
for any $F(\omega)\neq0$ in $L^{2}(S^{2},d\Omega)$. Decomposing
$\Psi(\omega)$ and $F(\omega)$ in spherical harmonics through (\ref{eq:a-ws-4}),
we get the following relation: $\langle\Psi_{\omega_{0},\chi,a}|F\rangle=\sum_{l\in\mathbb{N}}\sum_{|m|\leq l}\widehat{(\Psi_{\omega_{0},\chi,a})}_{lm}^{*}\widehat{F}_{lm}$,
with $\widehat{(\Psi_{\omega_{0},\chi,a})}_{lm}=\sum_{|n|\leq l}D_{mn}^{l}(\omega_{0},\chi)\widehat{(\Psi_{a})}_{ln}$
(see relation (\ref{eq:a-ws-6})). From the orthogonality relation
(\ref{eq:a-ws-5}) for the Wigner $D$-functions, the integral $I_{\Psi}^{F}$
finally reads:
\begin{equation}
I_{\Psi}^{F}=\sum_{l\in\mathbb{N}}C_{\Psi}^{l}\sum_{|m|\leq l}|\widehat{F}_{lm}|^{2}.\label{eq:a-cp-5}
\end{equation}
This proves that the condition (\ref{eq:a-cp-4}) implies
the wavelet admissibility condition (\ref{eq:ws-6}) on the sphere. The condition (\ref{eq:a-cp-4}) indeed requires that $C_{\Psi}^l$ be strictly positive and bounded for all $l$ by a strictly positive constant $c$, $0<C_{\Psi}^l<c$, while the wavelet admissibility condition (\ref{eq:ws-6}) reads $0<C_{\Psi}^l<\infty$. In the original group theoretic approach, the condition (\ref{eq:a-cp-4}) defines the admissibility, through the square-integrability of the considered group representation \cite{WSantoine2}. Our admissibility condition for the reconstruction of the signal is therefore slightly less restrictive.

\subsection{Establishment of the correspondence principle}

We can now prove the correspondence principle by showing that, if
$\psi(\vec{x})$ in $L^{2}(\mathbb{R}^{2},d^{2}\vec{x})$ satisfies
the reformulated wavelet admissibility condition (\ref{eq:a-cp-1}),
then the inverse stereographic projection $\Psi(\omega)=[\Pi^{-1}\psi](\omega)$
in $L^{2}(S^{2},d\Omega)$ satisfies the
condition (\ref{eq:a-cp-4}) \cite{WSbogdanova}, and therefore the admissibility condition (\ref{eq:ws-6}).

First, we establish the upper bound $I_{(\Pi^{-1}\psi)}^{F}<\infty$
for any $F(\omega)\neq0$ in $L^{2}(S^{2},d\Omega)$. The scalar product
of functions in $L^{2}(S^{2},d\Omega)$ is invariant under translation
by $\omega_{0}$, rotation by $\chi$. Considering a unitary operator
$\Pi$ between $L^{2}(S^{2},d\Omega)$ and $L^{2}(\mathbb{R}^{2},d^{2}\vec{x})$
implies by definition that the scalar product of functions is preserved
by the projection. Moreover, for a unitary, radial, and conformal
projection operator $\Pi$, the conjugation relation (\ref{eq:cp-3})
holds between the dilation operator $D(a)$ in $L^{2}(S^{2},d\Omega)$
on the sphere and the dilation operator $d(a)$ in $L^{2}(\mathbb{R}^{2},d^{2}\vec{x})$
on the plane. Indeed, the dilation operator $d(a)$ of functions in
$L^{2}(\mathbb{R}^{2},d^{2}\vec{x})$ on the plane defined by (\ref{eq:wp-3})
is also unitary, radial and conformal. The conjugate operator $\Pi^{-1}d\left(a\right)\Pi$
on functions in $L^{2}(S^{2},d\Omega)$ on the sphere therefore also
satisfies the same properties. Consequently, the uniqueness of the
unitary, radial and conformal dilation operator in $L^{2}(S^{2},d\Omega)$
on the sphere (see appendix \ref{sec:Wavelets-sphere-app}) implies
that the operator $\Pi^{-1}d\left(a\right)\Pi$ is identified with the
dilation $D(a)$ defined in (\ref{eq:ws-3}). We therefore get:
\begin{equation}
I_{(\Pi^{-1}\psi)}^{F}=\int_{SO(3)}d\rho\left[i_{(\Pi^{-1}\psi)}^{F}\left(\rho\right)\right],\label{eq:a-cp-6}
\end{equation}
with
\begin{equation}
i_{(\Pi^{-1}\psi)}^{F}\left(\rho\right)=\int_{0}^{+\infty}\frac{da}{a^{3}}\,|\langle d\left(a\right)\psi|\Pi R^{-1}\left(\rho\right)F\rangle|^{2}.\label{eq:a-cp-7}
\end{equation}
Similarly, from the invariance of the scalar product of functions
in $L^{2}(\mathbb{R}^{2},d^{2}\vec{x})$ under translation by $\vec{x}_{0}$
and rotation by $\chi$ we may write the upper bound of the wavelet
admissibility condition (\ref{eq:a-cp-1}) on $\psi(\vec{x})$ on
the plane as
\begin{equation}
I_{\psi}^{f}=\int_{0}^{2\pi}d\chi\int_{0}^{+\infty}\frac{da}{a^{3}}\int_{\mathbb{R}^{2}}d^{2}\vec{x}_{0}|\langle d\left(a\right)\psi|r^{-1}\left(\chi\right)t^{-1}
\left(\vec{x}_{0}\right)f\rangle|^{2}<\infty,\label{eq:a-cp-8}
\end{equation}
for any $f(\vec{x})\neq0$ in $L^{2}(\mathbb{R}^{2},d^{2}\vec{x})$.
By continuity of the integrand in the variables $\vec{x}_{0}$ and
$\chi$, this finally implies that
\begin{equation}
\int_{0}^{+\infty}\frac{da}{a^{3}}\,|\langle d\left(a\right)\psi|f\rangle|^{2}<\infty,\label{eq:a-cp-9}
\end{equation}
for any $f(\vec{x})\neq0$ in $L^{2}(\mathbb{R}^{2},d^{2}\vec{x})$.
Consequently, for any $F(\omega)\neq0$ in $L^{2}(S^{2},d\Omega)$,
the function $f(\vec{x})=[\Pi R^{-1}(\rho)F](\vec{x})$ is in $L^{2}(\mathbb{R}^{2},d^{2}\vec{x})$
and differs from zero, and we readily obtain that $i_{(\Pi^{-1}\psi)}^{F}(\rho)<\infty$,
for any $\rho\in SO(3)$. The compactness of the group $SO(3)$ finally
ensures that $I_{(\Pi^{-1}\psi)}^{F}<\infty$ for any $F(\omega)\neq0$
in $L^{2}(S^{2},d\Omega)$.

Notice that the choice of the measure of integration on scales $da/a^{3}$
is natural on the plane (see (\ref{eq:wp-5})), while \emph{a priori}
arbitrary on the sphere (see (\ref{eq:ws-5})). It is however
required to establish the correspondence principle between the wavelet formalisms
on the plane and on the sphere, as it clearly appears from the former
proof.

Second, the lower bound $0<I_{(\Pi^{-1}\psi)}^{F}$ for any $F(\omega)\neq0$
in $L^{2}(S^{2},d\Omega)$ remains to be established. In that regard,
we simply notice that the set of functions obtained by translation
by $\omega_{0}$, rotation by $\chi$, and dilation by $a$ of any
non-identically null function is dense in $L^{2}(S^{2},d\Omega)$.
The lower bound of the wavelet admissibility condition on the plane
(\ref{eq:a-cp-1}), $0<I_{\psi}^{f}$ for any $f(\vec{x})\neq0$ in
$L^{2}(\mathbb{R}^{2},d^{2}\vec{x})$, ensures that a wavelet on the
plane cannot be identically null. The unitarity of the stereographic
projection therefore implies that the inverse stereographic projection
$\Psi(\omega)=[\Pi^{-1}\psi](\omega)$ is also different from zero
in $L^{2}(S^{2},d\Omega)$. Consequently, the set of functions $\Psi_{\omega_{0},\chi,a}$
is dense in $L^{2}(S^{2},d\Omega)$, and considering any $F(\omega)\neq0$
in $L^{2}(S^{2},d\Omega)$, the scalar product $\langle\Psi_{\omega_{0},\chi,a}|F\rangle$
cannot be identically null in $\omega_{0}$, $\chi$, and $a$. Again,
the continuity of this function in all its arguments $\omega_{0}$,
$\chi$, and $a$ ensures that it is non-zero on a set of non-zero
measure in the corresponding Hilbert space. This strict positivity
of the integrand in $I_{(\Pi^{-1}\psi)}^{F}$ finally guarantees
that $0<I_{(\Pi^{-1}\psi)}^{F}$ for any $F(\omega)\neq0$ in $L^{2}(S^{2},d\Omega)$.
This completes the proof of the correspondence principle.

\subsection{Uniqueness of the stereographic projection operator}

Let us prove that the stereographic projection defined in (\ref{eq:cp-5})
is the unique unitary, radial, and conformal diffeomorphism between
the sphere and the plane. The projection operator $\Pi$ between functions
$G$ in $L^{2}(S^{2},d\Omega)$ on the sphere and $g$ in $L^{2}(\mathbb{R}^{2},d^{2}\vec{x})$
on the plane is generically expressed in terms the inverse of the
corresponding projection operator $\pi$ between points on the sphere
$S^{2}$ and on the plane $\mathbb{R}^{2}$, applied to the argument
of the function considered. The projection $\pi$ is a radial diffeomorphism,
that is a continuously differentiable bijection between $S^{2}$ and
$\mathbb{R}^{2}$, which only relates the radial variables $r$ on
the plane and $\theta$ on the sphere independently of $\varphi$,
and which leaves $\varphi$ invariant. In these terms, the operator
$\Pi$ and its inverse $\Pi^{-1}$ are given by the relations (\ref{eq:cp-5})
and (\ref{eq:cp-6}), with $\pi(\theta,\varphi)=(r(\theta),\varphi)$,
and its inverse $\pi^{-1}(r,\varphi)=(\theta(r),\varphi)$. The functions
$r(\theta)$ and its inverse $\theta(r)$, together with the function
$\mu(r)$, have to be fixed by the additional properties required.

First, the projection $\pi$ is a mapping between the
sphere and the plane, $\pi:\omega=(x^1,x^2)=(\theta,\varphi)\in S^2
\rightarrow\vec{x'}=(x'^1,x'^2)=(r,\varphi)\in\mathbb{R}^{2}$. It must be conformal
in order to preserve the measure of angles and directions, which are defined locally
in tangent plane at each point of $S^2$ and $\mathbb{R}^2$. This explicitly
means that the metric $g'_{ij}(r,\varphi)=(\partial x^k/\partial x'^i)
(\partial x^l/\partial x'^j)g_{kl}(\theta(r),\varphi)$ induced on the plane by
the projection from the sphere is conformally equivalent to the Euclidean
metric $g_{ij}(r,\varphi)=\textnormal{diag}(1,r^{2})$ on the plane, with $i,j,k,l\in\{1,2\}$.
The original metric on the sphere is the canonical metric
$g_{ij}(\theta,\varphi)=\textnormal{diag}(1,\sin^{2}\theta)$. The conformal equivalence reads
by definition: $g'_{ij}(r,\varphi)=e^{\phi(r)}g_{ij}(r,\varphi)$,
for some strictly positive conformal factor $e^{\phi(r)}$, that is
for a real function $\phi(r)$. This condition requires
that $r(\theta)$ is a linear function of $\tan\theta/2$. Let us
recall that we consider the plane tangent to the sphere at the North
pole ($z_{0}=1$). For the choice of coordinates $(r,\varphi)$ and
$(\theta,\varphi)$ defined in § \ref{sec:Correspondence-principle},
this gives uniquely:
\begin{equation}
r\left(\theta\right)=2\tan\frac{\theta}{2}.\label{eq:a-cp-10}
\end{equation}
This radial diffeomorphism $r(\theta)$ and its inverse $\theta(r)$
explicitly define the stereographic projection $\pi$ of points on
the sphere $S^{2}$ onto points on the plane $\mathbb{R}^{2}$, and
its inverse. By stereographic projection, the sphere without its South
pole is mapped onto the entire plane: $r(\theta):\theta\in[0,\pi[\rightarrow r\in[0,\infty[$.
A point $\omega=(\theta,\varphi)$ on the unit sphere is projected
onto a point $\vec{x}=(r,\varphi)$ on the tangent plane at the North
pole, co-linear with $\omega$ and the South pole (see Fig. \ref{cap:Stereographic-projection}).
The conformal factor is also explicitly given by the condition of
conformal mapping: $e^{\phi(r)/2}=(1+(r/2)^{2})^{-1}$, or equivalently
$e^{-\phi(r(\theta))/2}=1+\tan^{2}(\theta/2)$.

Second, the unitarity of the operator $\Pi$ on functions between
$L^{2}(S^{2},d\Omega)$ and $L^{2}(\mathbb{R}^{2},d^{2}\vec{x})$
identifies $\mu(r)$ with the conformal factor: $\mu(r)=e^{\phi(r)}$.
This identity simply relies on the relation between the invariant Euclidean
measure on the plane and the measure induced from the projection:
$r(\theta)dr(\theta)d\varphi=e^{-\phi(r(\theta))}\sin\theta d\theta d\varphi$.

In conclusion, the unique projection operator satisfying the properties
required to ensure a correspondence principle between the formalisms
of wavelets on the plane and on the sphere is therefore the stereographic
projection operator (\ref{eq:cp-5}).

\end{document}